\documentclass[pdflatex,sn-mathphys-num]{sn-jnl}

\usepackage{amsmath,amssymb}
\usepackage{graphicx}
\usepackage{xcolor}
\usepackage{booktabs}
\usepackage{multirow}
\usepackage{siunitx}
\usepackage{bm}
\usepackage{mathtools}

\newcommand{\gdir}{g_{\mathrm{dir}}}
\newcommand{\gada}{g_{\mathrm{mix}}}
\newcommand{\gstd}{g_{\mathrm{std}}}
\newcommand{\Eloc}{E_{\mathrm{loc}}}
\newcommand{\Eref}{E_{\mathrm{ref}}}
\newcommand{\ii}{\mathrm{i}}

\title[Low-variance phase-gradient estimators]{Low-variance estimators overcome the phase-gradient bottleneck in complex-valued neural quantum states}

\author[1,2]{Yi-Ran Xue}
\author*[1,4,5,6]{Rui Wang}
\email{rwang89@nju.edu.cn}
\author*[1,4,5]{Baigeng Wang}
\email{bgwang@nju.edu.cn}
\author*[1,2,3]{Chenan Wei}
\email{chenanwei@umass.edu}

\affil*[1]{National Laboratory of Solid State Microstructures and Department of Physics, Nanjing University, Nanjing 210093, China}
\affil[2]{Department of Physics, University of Massachusetts, Amherst, Massachusetts 01003, USA}
\affil[3]{A. Alikhanyan National Science Laboratory, Br. Alikhanian 2, Yerevan 0036, Armenia}
\affil[4]{Collaborative Innovation Center of Advanced Microstructures, Nanjing University, Nanjing 210093, China}
\affil[5]{Jiangsu Physical Science Research Center}
\affil[6]{Hefei National Laboratory, Hefei 230088, People's Republic of China }

\abstract{
Complex neural quantum states are difficult to optimize when their wavefunction phase carries gauge, chiral, fermionic, or topological structure. We show that the major failure mode is not only ansatz expressivity, but the Monte Carlo estimator used to learn this phase. For separated amplitude-phase states, differentiating the local energy at fixed samples gives a different unbiased estimator of the same variational Monte Carlo phase force, without changing the objective. We further extend the construction to coupled two-head networks by keeping the amplitude-gradient contribution and applying the direct derivative only to the phase path. An adaptive minimum-variance mixture interpolates between standard and direct estimators during training. Across flux ladders, chiral chains, two-dimensional flux cylinders, an interacting fermion ladder, shared-network controls, and a fractional quantum Hall benchmark, the resulting estimators reduce phase-gradient variance, suppress seed failures, and often move multi-percent standard-gradient plateaus to sub-percent accuracy.
}


\begin{document}

\maketitle

\section{Introduction}

Neural-network quantum states (NQS) have evolved from a proof-of-principle variational ansatz into a computational platform for quantum many-body science. In lattice models, neural wavefunctions now support symmetry resolution, autoregressive sampling, recurrent and transformer architectures, and large-scale software frameworks \cite{carleo2017solving,gao2017efficient,choo2018symmetries,sharir2020deep,hibatallah2020recurrent,nomura2021dirac,vicentini2022netket,medvidovic2024neural,lange2024review}. In quantum chemistry, materials science and related continuum problems, fermionic neural wavefunctions and neural-network quantum Monte Carlo have reached molecular and solid-state systems, and recent work has emphasized transferability, spin symmetry, efficient Laplacians, and foundation-style neural wavefunctions for correlated electrons \cite{han2019solving,pfau2020abinitio,hermann2020deep,li2022solids,scherbela2022solving,li2024forwardlaplacian,li2024spinsymmetry,gerard2025transferable,tang2025deep,luo2023pairing,teng2025solving,qian2025describing,zaklama2025attention,zaklama2026large,nazaryan2026qernel}. This progress has made a central computational question unavoidable: when a neural wavefunction fails on a difficult quantum state, is the limiting factor the expressivity of the ansatz, or the stochastic algorithm used to optimize it?

The difficult states in this question often have a nontrivial configuration-dependent wavefunction phase. For a basis configuration $x$, an amplitude--phase neural wavefunction has the form $\psi(x)=\exp[u(x)+\ii\phi(x)]$. The log amplitude $u(x)$ sets the sampling weight $|\psi(x)|^2$, while the phase function $\phi(x)$ determines relative phases between configurations. In VMC this phase enters through local-energy ratios $\psi(y)/\psi(x)$: the relative phase $\phi(y)-\phi(x)$ between configurations connected by the Hamiltonian controls the off-diagonal interference terms sampled by the optimizer. Nontrivial phase structure appears in many of the systems that motivate complex NQS: fermionic antisymmetry underlies the sign problem in generic quantum Monte Carlo \cite{troyer2005computational}; frustrated magnets evade simple Marshall sign rules \cite{marshall1955antiferromagnetism}, and previous NQS studies have identified sign structure as a source of poor generalization, rugged landscapes, and optimizer failure \cite{westerhout2020generalization,szabo2020neural,bukov2021learning,chen2022neural}; gauge fields and twisted boundary conditions introduce Peierls phases and Hofstadter-type interference patterns \cite{orignac2001meissner,atala2014observation,hugel2014chiral,ledinauskas2025universal,doschl2025hofstadter}; time-reversal-symmetry-breaking interactions generate complex spin wavefunctions \cite{wei2024unveiling}; quantum Hall and other topological states carry intrinsically complex many-body structure \cite{teng2025solving,qian2025describing}; and unitary real-time dynamics makes the wavefunction complex by construction \cite{schmitt2020quantum}. In these settings, accurate amplitudes alone are not sufficient: the reliability of the phase-gradient sector directly affects the variational energy reached by optimization.

Most NQS strategies for hard states improve what the ansatz can express or how strongly physics is built into it: larger networks, symmetries, backflow, attention, pretraining, or more aggressive phase-network learning rates \cite{chen2024empowering,zhang2023transformer,ou2025improving}. These developments are essential, but they leave a separate statistical question: even when an ansatz can support the needed wavefunction phase, is the Monte Carlo gradient used to optimize its phase parameters stable enough? Recent work on VMC importance sampling has begun to address this estimator side by adapting the sampling distribution to reduce high-variance gradient estimates \cite{misery2025looking}. Here we take a different but aligned route: the sampling law is kept fixed, and only the phase-gradient random variable is changed. In the 100-site flux ladder, a compact width-128, depth-2 ansatz trained with the direct estimator converges below one percent of the DMRG reference, whereas standard-gradient networks widened or deepened under the same optimizer settings remain at much larger errors. This comparison is not a claim that larger networks are intrinsically worse after separate retuning; it shows that increasing capacity is a less direct remedy than reducing the noise in the phase-force estimator. Increasing the phase-network learning rate partly improves the standard estimator but only within a narrow range, after which it plateaus or destabilizes. This tuning sensitivity is itself part of the failure mode: when the raw gradient estimator has a high variance floor, scalar step-size tuning rescales signal and noise together rather than changing the random variable being estimated. The bottleneck can therefore be statistical, residing in the Monte Carlo estimator of the phase gradient rather than in ansatz capacity alone.

The mechanism is simple but consequential. The conventional VMC gradient treats amplitude and phase through one energy-gradient covariance, often preconditioned by stochastic reconfiguration or the quantum natural gradient \cite{sorella1998green,becca2017variational,stokes2020quantum}. For a separated amplitude--phase ansatz, however, the phase force is a covariance between $\mathrm{Im}\,\Eloc$ and the phase score. It is therefore a score-function estimator in the same sense as REINFORCE gradients \cite{williams1992simple,mohamed2020monte}. In broader Monte Carlo gradient estimation, score-function estimators are often contrasted with pathwise or reparameterized derivatives, which can have much lower variance when the sampling distribution permits them \cite{kingma2014auto,mohamed2020monte}. The standard VMC phase estimator has the correct expectation, but its signal-to-noise ratio can collapse pre-asymptotically: the covariance can be small through cancellation even when the sample-wise fluctuations of $\mathrm{Im}\,\Eloc$ remain large. The zero-variance principle only guarantees that this noise disappears at an exact eigenstate \cite{assaraf1999zero}; practical optimization must reach that limit through a noisy transient regime.

Here we use the structure of the wavefunction-phase sector to construct a lower-variance route to the same force. In a separated amplitude--phase parameterization, the sampling distribution $|\psi|^2$ depends on the log amplitude but not on $\phi(x)$. One may therefore differentiate the local energy itself at fixed sampled configurations. The resulting direct estimator $\gdir$ is unbiased for the same phase force as the standard score estimator $\gstd$, but it is a different Monte Carlo random variable with different variance. It requires no knowledge of the model beyond the connected configurations already used to evaluate the local energy. Because neither endpoint has uniformly lower variance along a training trajectory, we also introduce an adaptive mixture $\gada=(1-\lambda^\star)\gstd+\lambda^\star\gdir$. The coefficient is the clipped minimum-variance coefficient for the two raw unbiased phase-gradient estimators. In population, this construction identifies the minimum-variance convex interpolation of the two raw estimators and gives an explicit sensitivity identity for coefficient-estimation error; in finite-batch preconditioned VMC it should be read as a variance-aware estimator-combination rule, which we monitor directly through logged variance ratios.

We test the resulting estimator-design narrative on flux ladders, FQHE, chiral XXX chains, two-dimensional flux squares, an interacting spinless-fermion ladder, exact-diagonalization controls, and zero-flux real-valued controls. The theorem and the main lattice benchmarks use separated amplitude and phase networks; later we state the coupled two-head generalization and use single-network calculations as supporting transfer tests. The main benchmark is a 100-site flux ladder with DMRG references \cite{white1992density,schollwock2011density}. At $\Phi=0.3\pi$, $\gdir$ reaches a median tail-window error of $0.83\%$ and the adaptive mixture $0.57\%$, while standard baselines with tuned phase-network learning rates remain in the $3.4$--$5.1\%$ range and the unscaled standard estimator near $11\%$. At $\Phi=0.2\pi$, the adaptive mixture reaches $0.52\%$, while tuned standard baselines remain at $2.7$--$4.7\%$. Starting from DeepHall, the adaptive-mixture optimization reaches an energy below the second-Landau-level-projected exact-diagonalization reference, consistent with accessing Landau-level-mixing correlations beyond that fixed-level projection. The gain transfers to chiral XXX chains, where the standard-to-direct mean-error ratio is $8.9\times$ at $N=20$, $\alpha=1.0$ and $4.85\times$ at $N=100$, to an $8\times8$ flux square with high-bond-dimension finite-MPS DMRG references, and to the fermionic flux-ladder control in Extended Data Fig.~\ref{fig:ed_fermion}. Variance diagnostics, flux controls, and seed-resolved statistics all support the same conclusion: optimization of the complex wavefunction phase in NQS is not only an architecture problem, but a Monte Carlo gradient-estimator problem.

\section{Results}
The benchmark suite spans distinct physical origins of a nontrivial wavefunction phase $\phi(x)$. The flux ladder is a quasi-one-dimensional gauge-field model, the chiral XXX chain introduces time-reversal-symmetry breaking through scalar chirality, the two-dimensional flux square realizes a Hofstadter-type lattice gauge field, the interacting spinless-fermion ladder combines Peierls phases with fermionic signs and density interactions, and the fractional quantum Hall benchmark probes continuum topological states with Landau-level mixing. Zero-flux and real frustrated controls remove or reduce the complex phase structure. The separated-network lattice calculations are the main validation of the estimator identity; the single-network FQHE and shared-trunk ladder calculations test how the same estimator idea transfers beyond that clean setting. Together these models provide controlled variations in the origin and strength of wavefunction-phase structure while retaining reference energies from exact diagonalization, projected exact diagonalization, or DMRG. The Hamiltonians and benchmark sectors are collected in Methods.

\subsection{Direct/adaptive-mixture phase gradients in a flux ladder}

We first use the flux ladder as a controlled setting in which the target wavefunction is complex but the reference energy is accurately accessible. The model is a two-leg spin-$1/2$ ladder with $L=50$ rungs ($100$ sites), fixed magnetization, and a plaquette flux $\Phi$; it is closely related to gauge-field ladder realizations of Meissner and vortex physics \cite{orignac2001meissner,hugel2014chiral,atala2014observation}. At finite flux, Peierls phases on the leg exchanges generate chiral currents and a tunable interference pattern in the ground-state wavefunction, making the ladder a compact benchmark for wavefunction-phase optimization. The neural ansatz uses separated real networks for log-amplitude and phase. Unless stated otherwise, all comparisons share the same sampler, MLP architecture, stochastic-reconfiguration/minSR pipeline, clipping, damping, and learning-rate schedule; only the estimator of the phase-force vector is changed.

\begin{figure}[t]
\centering
\includegraphics[width=\linewidth]{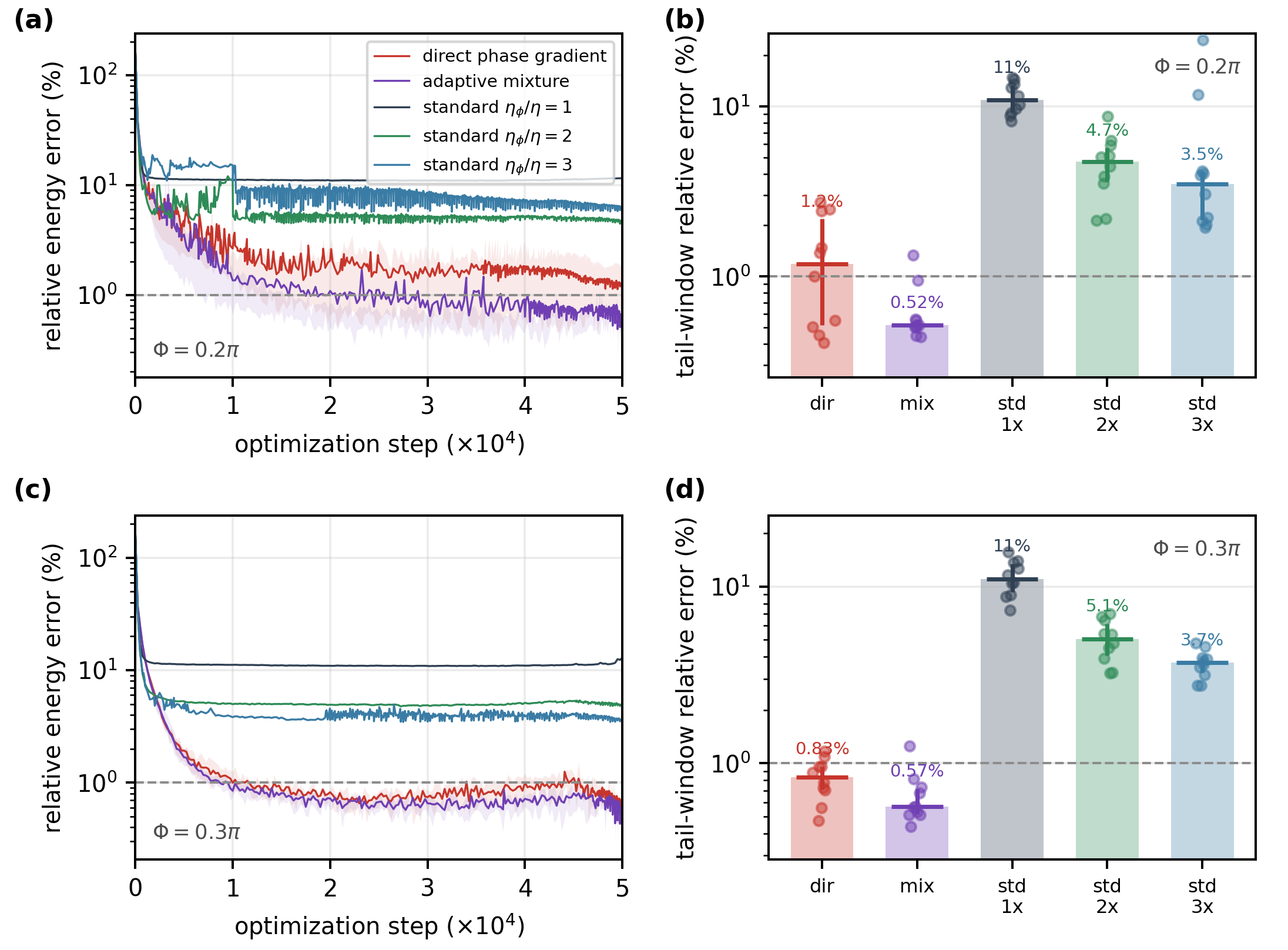}
\caption{\textbf{Direct and adaptive-mixture phase gradients on the 100-site flux ladder.}
All panels use the $L=50$ flux ladder ($100$ sites, half filling), an MLP with width $128$ and depth $2$, batch size $1024$, $48$ MCMC sweeps per optimization step, base learning rate $\eta=0.03$, and ten seeds per estimator.
Each row is one flux: $\Phi=0.2\pi$ (top, DMRG $\Eref=-44.826$) and $\Phi=0.3\pi$ (bottom, DMRG $\Eref=-43.303$). (Left column) relative-error training curves; solid lines show means, dotted lines medians, and shaded bands interquartile ranges. (Right column) tail-window relative-error statistics for the direct, adaptive-mixture, and standard estimators including the phase-learning-rate multipliers; bars show medians, whiskers interquartile ranges, and points individual seeds.}
\label{fig:flux}
\end{figure}

Figure~\ref{fig:flux} compares the standard score phase gradient with the direct local-energy derivative and the adaptive mixture. At $\Phi=0.3\pi$, the direct estimator reaches a median tail-window relative error of $0.83\%$ (mean $0.83\%$, range $0.47$--$1.17\%$ over ten seeds), and the adaptive mixture reaches $0.57\%$ (range $0.44$--$1.26\%$). The unscaled standard estimator remains near $11.1\%$ and shows a bimodal seed distribution. We also sweep the standard phase-network learning-rate multiplier to separate estimator variance from a simple step-size mismatch: if the gap were only a step-size effect, a larger phase-network step would close it. Across $\eta_\phi/\eta\in\{2,3,4,5\}$, the standard estimator still plateaus at $3.4$--$5.1\%$ (Extended Data Fig.~\ref{fig:ed_phaselr}). At $\Phi=0.2\pi$, the same ordering holds. The adaptive mixture gives the lowest median error, $0.52\%$ (range $0.44$--$1.33\%$), the direct estimator gives $1.2\%$, the unscaled standard estimator remains near $10.9\%$, and the tuned standard baselines remain in the $2.7$--$4.7\%$ range. Thus, tuning the score estimator helps only partially and introduces an additional search cost, whereas replacing the phase-gradient random variable moves the calculation from multi-percent plateaus to sub-percent accuracy in the same optimization pipeline.

\subsection{Estimator choice is more efficient than increasing capacity}

\begin{figure}[t]
\centering
\includegraphics[width=\linewidth]{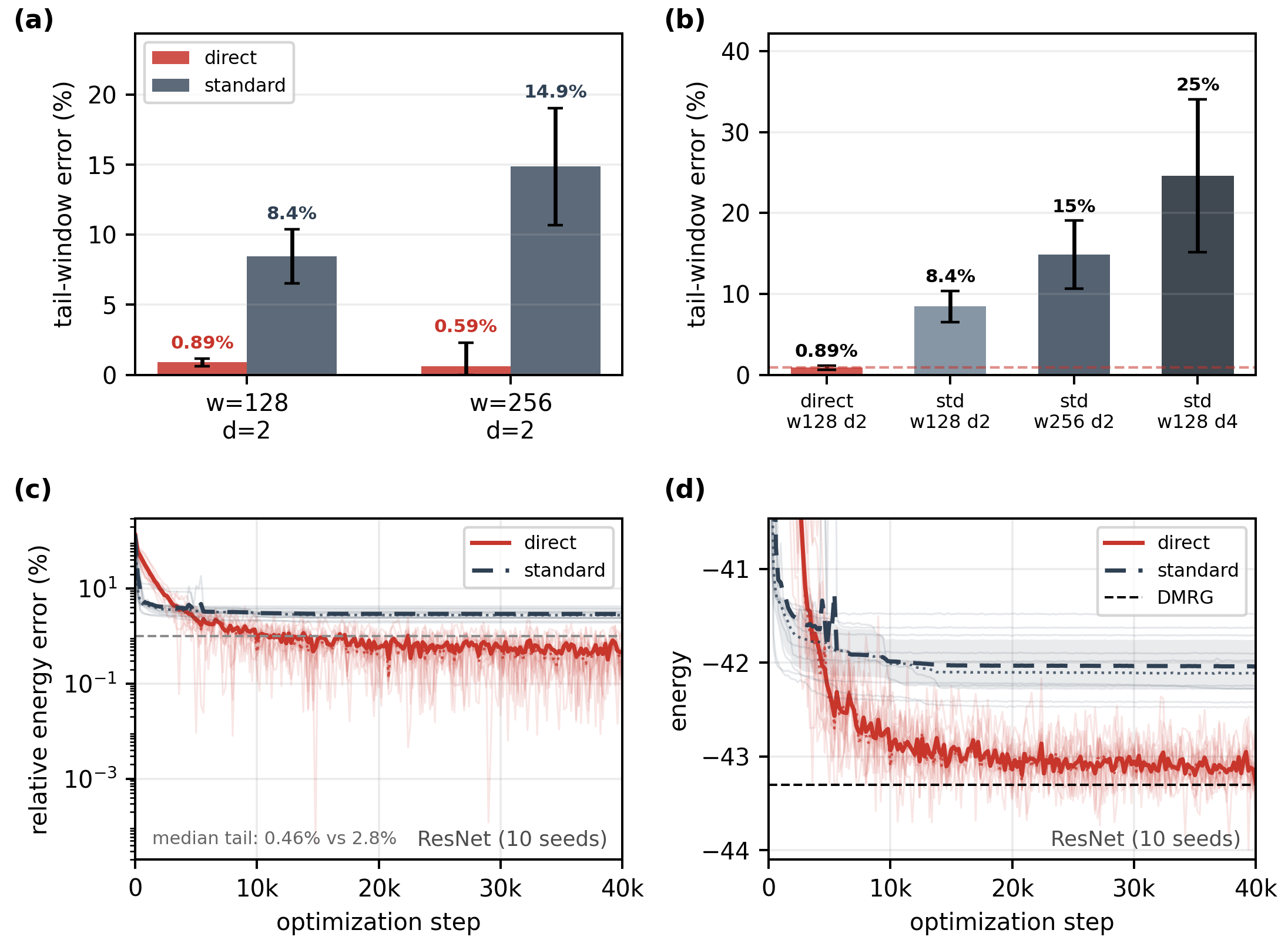}
\caption{\textbf{Estimator choice beats capacity scaling under matched training.}
The benchmark is the $L=50$ ($100$-site) flux ladder at $\Phi=0.3\pi$, with the same optimizer settings as Fig.~\ref{fig:flux} for every architecture, except that the ResNet comparison (panels c,d) uses batch size $512$ for the standard-gradient runs and $256$ for the direct runs (an asymmetry that favors the standard baseline).
(a) Width sweep: tail-window error of the direct and standard phase-gradient estimators at width $128$ and width $256$ (depth $2$).
(b) Architecture comparison: the shallow direct network (width $128$, depth $2$) against standard-gradient MLPs of width $128$ depth $2$, width $256$ depth $2$, and width $128$ depth $4$; under these matched optimizer settings, increasing standard-gradient capacity does not close the estimator gap.
(c,d) ResNet comparison over ten seeds (thin lines, individual seeds; solid, mean; dotted, median; shaded band, interquartile range): relative-error (c) and energy (d) convergence, with the direct estimator descending past the standard plateau to a median tail-window error of $0.46\%$ versus $2.75\%$ for the standard gradient.}
\label{fig:capacity}
\end{figure}

A natural interpretation of poor wavefunction-phase optimization is that the network is too small. We therefore compare the estimator change with standard-gradient capacity changes under matched optimizer hyperparameters. In Fig.~\ref{fig:capacity}, the direct estimator with the shallow width-128, depth-2 MLP reaches a median tail-window error below $1\%$ ($0.89\%$). The standard estimator does not improve under the same training recipe: the width-128, depth-2 standard MLP gives $8.4\%$, the width-256, depth-2 MLP gives $14.9\%$, and the width-128, depth-4 MLP gives $24.6\%$. We do not use this sweep to argue that larger standard-gradient architectures cannot be rescued by their own tuning protocols. The point is different: a good phase-gradient estimator makes a compact ansatz efficient. It reduces the need to compensate for estimator noise by enlarging the network and reopening a separate architecture-by-architecture hyperparameter search.

The same conclusion survives an architecture change. A translation-equivariant convolutional ResNet with spatial sum-pooling ($24$ channels, depth $6$) again shows a clear estimator gap: the direct estimator continues to descend after the standard runs have plateaued, reaching a median tail-window error of $0.46\%$ (range $0.27$--$1.03\%$) versus $2.75\%$ (range $1.92$--$4.21\%$) for the standard gradient. The ResNet comparison is conservative in favor of the baseline because the standard-gradient runs use twice the batch size of the direct runs ($512$ versus $256$). The practical message is that architecture and estimator are distinct levers; in these tests, changing the estimator is the more efficient way to make the phase sector train reliably.

\subsection{Controls identify a phase-specific variance mechanism}

\begin{figure}[t]
\centering
\includegraphics[width=\linewidth]{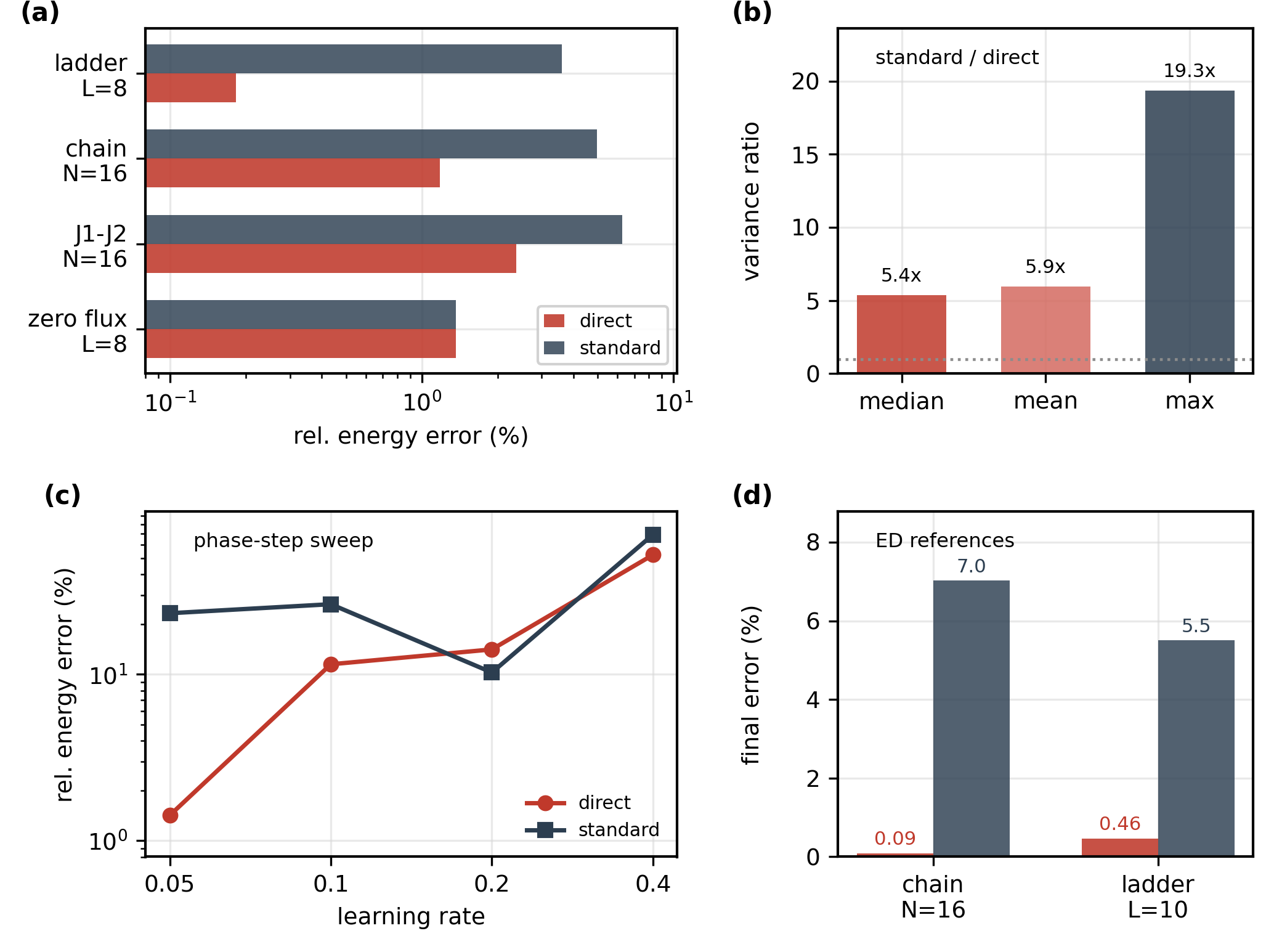}
\caption{\textbf{Mechanism and controls.}
(a) Mean relative-error controls for a flux ladder with $L=8$, a flux chain with $N=16$, a frustrated $J_1$--$J_2$ flux chain with $N=16$, and a zero-flux ladder control; the large ratios in complex models come mainly from fewer bad-basin seeds for the direct estimator.
(b) Standard/direct phase-gradient variance ratios on a representative complex-flux trajectory, summarized by median, mean, and maximum ratio.
(c) Phase-learning-rate sweep for estimators over the displayed learning-rate multipliers.
(d) Exact-diagonalization controls for a chain with $N=16$ and a ladder with $L=10$.
The advantage appears as improved robustness in complex flux models and disappears in the zero-flux real control.}
\label{fig:mechanism}
\end{figure}

The estimator advantage should appear only when the phase sector is statistically difficult. Figure~\ref{fig:mechanism} tests this prediction. The small flux ladder and flux chain are finite-size versions of the Peierls-phase setting, where complex hopping or exchange phases produce a controlled wavefunction phase. The frustrated $J_1$--$J_2$ chain introduces nontrivial sign structure through competing exchanges, providing a useful comparison between discrete sign structure and irreducibly complex phases. In the clean zero-flux bipartite control, the phase structure is trivial, and the standard and direct estimators have essentially identical mean errors. These controls separate the method from a generic optimization improvement: the gain follows the presence of a complex phase.

The same figure also gives two diagnostics for the proposed mechanism. First, on a representative complex-flux trajectory, the standard-to-direct phase-gradient trace-variance ratio has median $5.4\times$ and maximum $19.3\times$. Second, exact-diagonalization controls on small chains and ladders reproduce the large standard/direct gaps without relying on DMRG reference uncertainty. Together with the learning-rate sweep, these results support the variance-limited picture: multiplying the noisy phase update can partly improve early descent, but it does not remove the raw estimator variance and eventually destabilizes.

\subsection{Adaptive mixing makes estimator choice variance-aware}

The direct estimator is not expected to dominate universally. Near an exact eigenstate, the standard estimator inherits the usual zero-variance property, whereas the direct estimator need not be zero-variance componentwise. The adaptive mixture is designed for this changing regime. It forms a convex combination of the standard and direct phase estimators using a trace minimum-variance coefficient estimated on the pre-update ensemble, with convex clipping of the reported coefficient to $0\le\lambda^\star\le1$.

\begin{figure}[t]
\centering
\includegraphics[width=\linewidth]{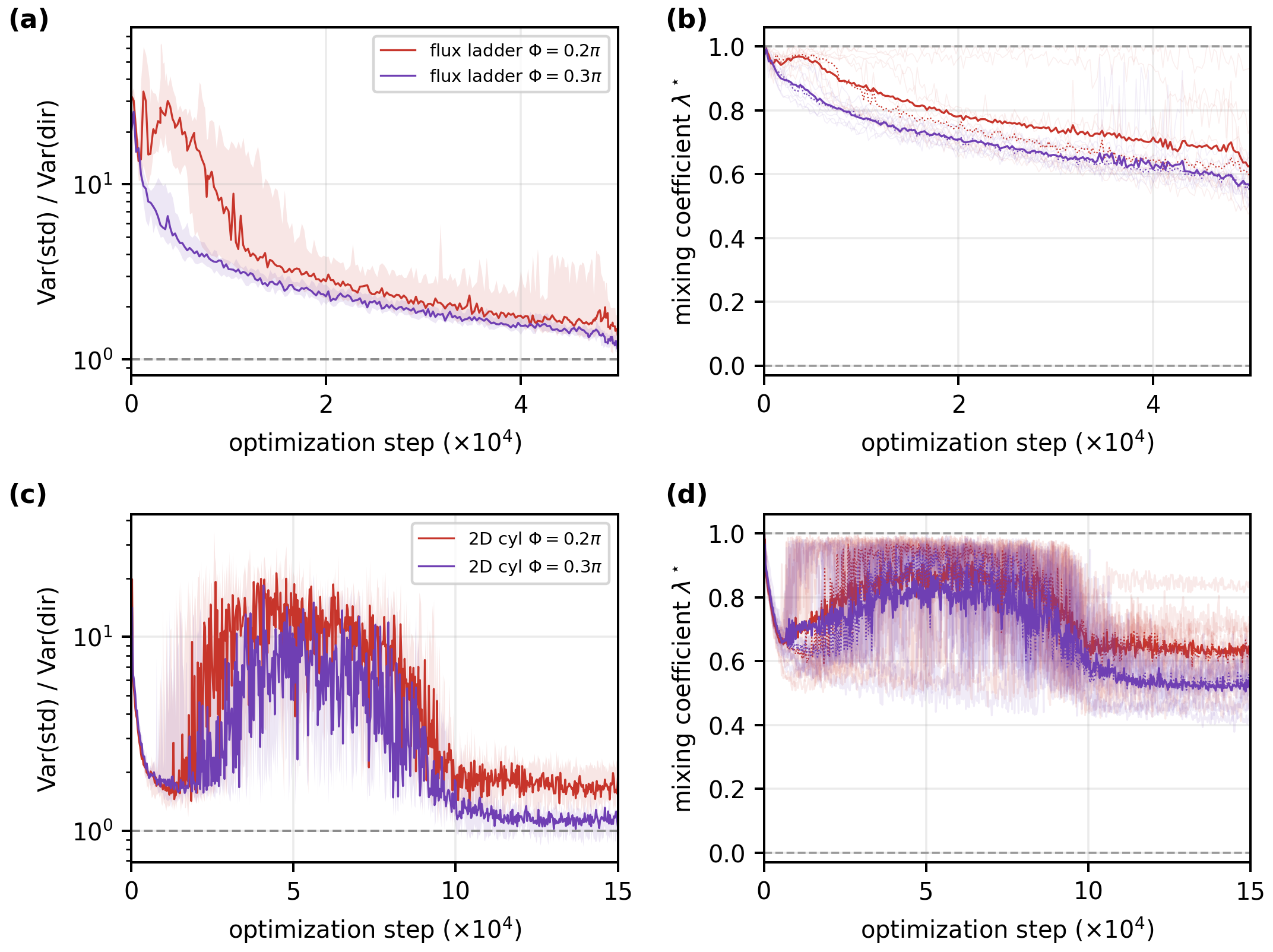}
\caption{\textbf{Gradient-variance reduction and adaptive mixing.}
Logged phase-gradient variance ratios and mixing coefficient $\lambda^\star$ along training, for the $L=50$ flux ladder at $\Phi=0.2\pi$ and $\Phi=0.3\pi$ and for the 2D cylinder ($\Phi=0.2\pi$ and $0.3\pi$, $J_z=0$). Solid curves show $\mathrm{Var}(\mathrm{standard})/\mathrm{Var}(\mathrm{direct})$; the mixing-coefficient panels show means (solid), medians (dotted), individual seeds (thin), and interquartile bands.}
\label{fig:variance}
\end{figure}

Figure~\ref{fig:variance} logs the variance ratios and the mixing coefficient during training. On both flux-ladder points and on the 2D flux square, the standard estimator has larger and more persistent phase-gradient variance over much of the trajectory. The adaptive mixture tracks a lower-variance interpolation between the standard and direct estimators over much of training without imposing a fixed interpolation. The learned coefficient starts close to the direct endpoint and drifts away from $\lambda^\star=1$ as training proceeds, matching the Methods prediction that the population-optimal raw-gradient mixture returns toward the standard estimator as the zero-variance regime is approached. These diagnostics are important because the theory is a statement about raw gradient estimators, whereas the production calculation also contains finite-batch coefficient estimates, clipping, and stochastic-reconfiguration preconditioning. The logged behavior shows that the complete pipeline uses the variance information in the intended direction.

\subsection{Continuous-space generality: Landau-level mixing in the fractional quantum Hall effect}

\begin{figure}[t]
\centering
\includegraphics[width=\linewidth]{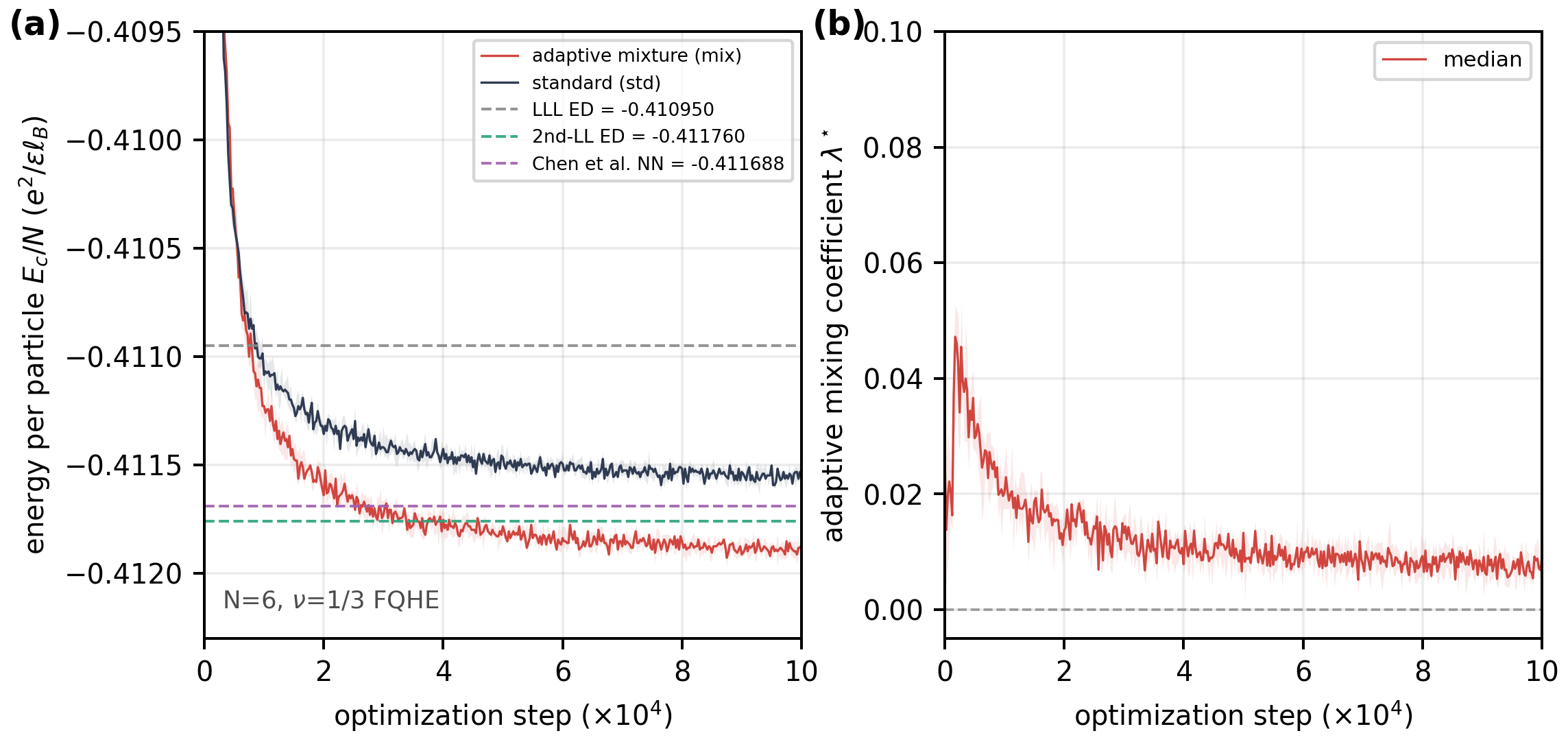}
\caption{\textbf{Generality to a continuous-space fractional quantum Hall benchmark (DeepHall).}
Single-network neural quantum state built on the transform for $N=6$ electrons at filling $\nu=1/3$ on the Haldane sphere, trained with the adaptive-mixture (mix) and standard (std) phase-gradient estimators, ten seeds per estimator. (a) Corrected energy per particle versus optimization step (median over seeds; shaded band, interquartile range; per-seed curves rolling-median smoothed). Dashed lines mark the lowest-Landau-level exact-diagonalization energy ($-0.410950$), the two-Landau-level ED energy ($-0.411760$, gap $=1$), and the published DeepHall neural baseline ($-0.411688$). The adaptive mixture descends below the two-Landau-level ED value, consistent with beyond-two-Landau-level mixing, while the standard estimator stays above it; a high-statistics MCMC evaluation at fixed final parameters gives $E_c/N=-0.411898$ for the adaptive mixture and $E_c/N=-0.411555$ for the standard estimator. (b) Adaptive mixing coefficient $\lambda^\star$ along training, shown on an expanded vertical scale.}
\label{fig:deephall}
\end{figure}

The estimator-design principle is not specific to lattice spin models. It also applies when a continuous-coordinate neural state learns a complex phase. We use the fractional quantum Hall effect (FQHE) as a test of this transfer. The ground state is not confined to a single Landau level (LL); it admixes higher-LL components, and capturing this \emph{Landau-level mixing} is precisely where a flexible neural ansatz can go beyond a fixed-LL exact-diagonalization (ED) treatment. We therefore study $N=6$ electrons at filling $\nu=1/3$ on the Haldane sphere with a DeepHall ansatz, a single complex network in the FermiNet family, trained with the standard and adaptive-mixture phase-gradient estimators under otherwise identical settings, ten seeds per estimator (Fig.~\ref{fig:deephall}). This benchmark is intentionally presented as a transfer test beyond the separated-network lattice setting developed above.

The adaptive-mixture estimator drives the energy \emph{below} the two-Landau-level ED benchmark---the ground-state energy obtained by exact diagonalization restricted to the lowest two LLs---whereas the standard phase-gradient estimator plateaus above it. In the high-precision frozen evaluation, the adaptive mixture reaches $E/N=-0.411898\,e^2/\epsilon\ell_B$, below the two-LL ED value $-0.411760$ (LL gap$=1$) by $1.4\times10^{-4}$, while the standard estimator settles at $-0.411555$, above the two-LL ED value by $2.0\times10^{-4}$. The crossing of the two-LL reference is consistent with the adaptive-mixture network accessing beyond-two-LL mixing correlations, the regime in which a fixed two-LL projection remains variationally constrained. The adaptive mixture also improves on the previously reported DeepHall neural-network energy ($-0.411688$), giving the lowest value among the compared methods on this benchmark. The learned mixing coefficient in Fig.~\ref{fig:deephall}b is smaller than in the lattice flux ladders but remains nonzero on the expanded vertical scale, so the continuum result should be read as a sensitive variance-aware correction to the FermiNet-class optimization rather than as a wholesale replacement of the standard gradient. The FQHE result therefore strengthens the main message: estimator choice can decide whether a complex neural state resolves a small but physically meaningful phase-correlation energy.

\subsection{The estimator advantage persists across lattice phase structures}

We next test whether the advantage is tied to a particular ladder Hamiltonian. The chiral XXX benchmark is a time-reversal-breaking spin chain with Heisenberg exchange supplemented by a scalar-chirality interaction, motivated by recent finite-size and conformal-field-theory studies of chiral spin states \cite{wei2024unveiling}. The coupling $\alpha$ tunes the strength of chirality and therefore the amount of complex wavefunction structure generated without Peierls hopping phases. Extended Data Fig.~\ref{fig:ed_chiral} shows that the direct phase gradient remains better across chirality strength, system size, and optimizer setting. At $N=20$, $\alpha=1.0$, the standard-to-direct mean-error ratio is $8.9\times$; at $N=100$, the ratio remains $4.85\times$. The transfer from Peierls flux to scalar chirality is a useful stress test because the source of the complex phase is physically different while the estimator structure is the same.

\begin{figure}[t]
\centering
\includegraphics[width=\linewidth]{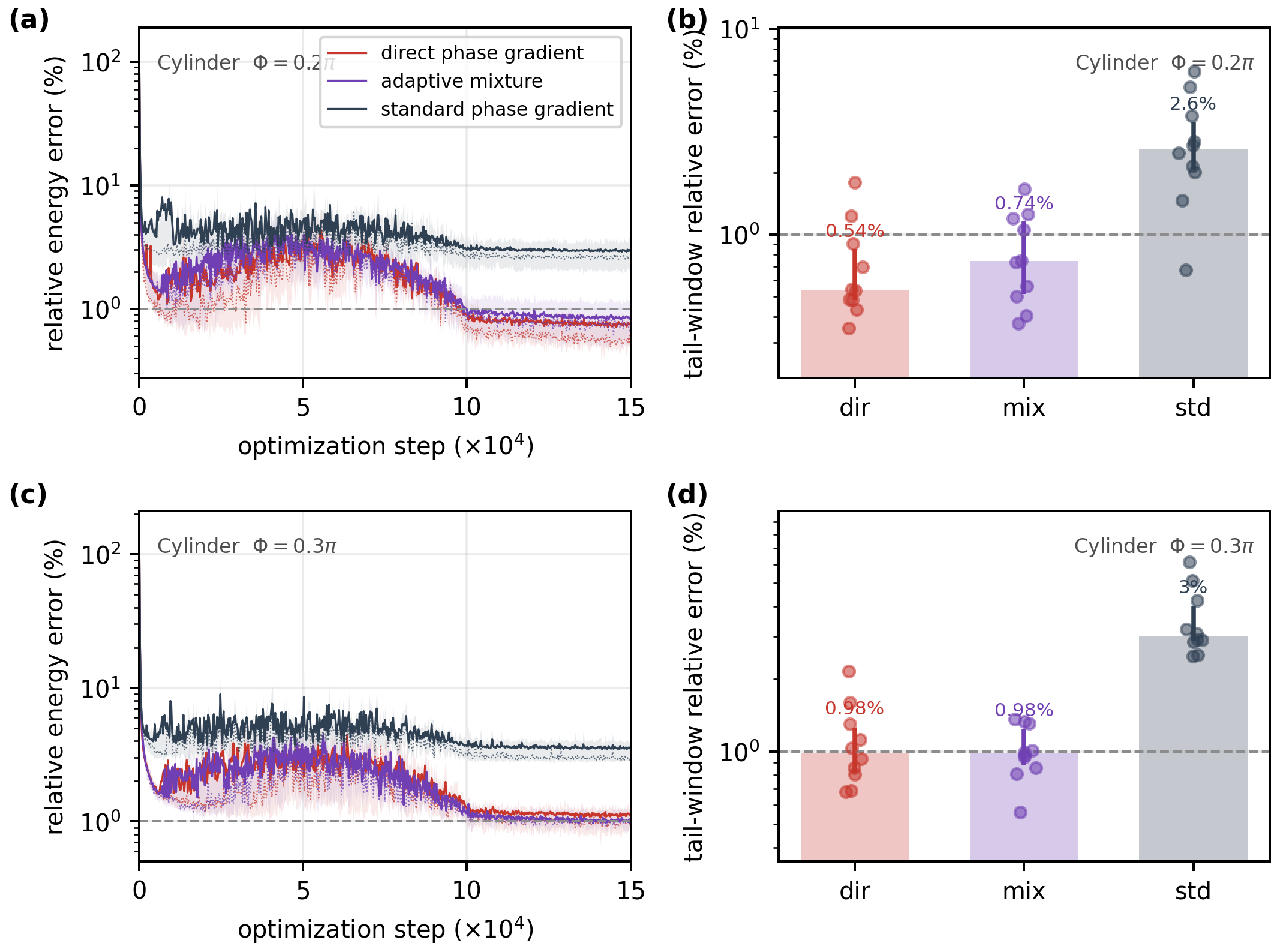}
\caption{\textbf{Generality on a two-dimensional $8\times8$ flux square.}
Relative-energy-error training trajectories (left column) and tail-window relative-error statistics (right column; bars, median; whisker, interquartile range; points, individual seeds) for the direct, adaptive-mixture, and standard estimators at $J_z=0$, on an $8\times8$ cylinder (periodic $x$, open $y$) with a uniform Peierls flux per plaquette ($\Phi=0.2\pi$, top row; $\Phi=0.3\pi$, bottom row), ten seeds per estimator. Reference energies are high-bond-dimension finite-MPS DMRG values ($\chi$ up to 768; Methods). At fixed cylinder geometry the standard estimator is both higher and more broadly dispersed than the direct and adaptive-mixture estimators, and its error grows with flux; the same ordering at the interacting point $J_z=0.5$ is shown in Extended Data Fig.~\ref{fig:ed_2d}.}
\label{fig:square2d}
\end{figure}

In addition, we study an $8\times8$ square lattice with a uniform Peierls flux per plaquette (Fig.~\ref{fig:square2d}), using high-bond-dimension finite-MPS DMRG references. This two-dimensional flux square is a spin analogue of a Hofstadter-type lattice gauge-field problem: the Landau-gauge Peierls phases make the off-diagonal interference pattern genuinely two-dimensional, while the $J_z$ term controls the strength of density--density interactions in the spin representation. The main figure shows $J_z=0$, where the flux-induced complex phase is the only source of sign structure. On a cylinder (periodic $x$, open $y$) at $\Phi=0.2\pi$, the direct estimator reaches $0.54\%$ median tail-window error and the adaptive mixture $0.74\%$, while the standard estimator plateaus at $2.61\%$ with a broader seed distribution. At $\Phi=0.3\pi$ the direct estimator and the adaptive mixture both reach $0.98\%$, compared with $3.00\%$ for the standard estimator. Adding an interaction $J_z=0.5$ leaves the ordering intact: at $\Phi=0.2\pi$ the adaptive mixture reaches $2.37\%$ versus $6.59\%$ standard, and at $\Phi=0.3\pi$ the direct estimator reaches $3.65\%$ versus $7.57\%$ standard (Extended Data Fig.~\ref{fig:ed_2d}). Across the four two-dimensional cells, the standard-to-lower-error estimator ratio spans $2.1$--$4.8\times$. The gain is again mainly reliability: the standard estimator often converges to a negative energy, but its seed distribution is higher and wider. Extended Data Fig.~\ref{fig:ed_fermion} shows the same qualitative ordering on a 100-site interacting spinless-fermion flux ladder with a high-bond-dimension finite-MPS DMRG reference. This fermionic benchmark combines Peierls phases, density interactions, and particle-exchange signs after Jordan--Wigner mapping, indicating that the estimator effect is not restricted to spin-only benchmarks.

Finally, we include a deliberately smaller shared-parameter test in Extended Data Fig.~\ref{fig:ed_singlenet}. The main proof and main lattice figures use separate amplitude and phase parameter blocks. The shared-trunk control instead uses one trunk with two output heads and the coupled-force construction described in Methods: the amplitude/Pulay score term is retained, while the direct summand differentiates only the phase path with the amplitude path stopped. The direct and adaptive-mixture estimators remain better than the standard estimator in this setting, so the observed advantage is not merely an artifact of splitting amplitude and phase into two independent networks. Because this shared-network evidence is narrower than the separated-network benchmark suite, we treat it as a transfer check rather than the main validation of the theory.

\section{Discussion}

The central message is that complex-phase NQS optimization can be a gradient-estimator problem. Although the standard VMC phase force is unbiased, its score-function covariance signal can become small through cancellation while fluctuations of $\mathrm{Im}\,\Eloc$ remain large. In this regime, replacing the estimator can be more efficient than increasing capacity: on the flux ladder, wider or deeper standard-gradient networks did not close the gap at matched settings, whereas the compact direct-gradient ansatz reached sub-percent error. The chiral-chain and two-dimensional flux-square benchmarks show that the effect is not ladder-specific.

The estimator identity also fixes the scope. For strictly separated amplitude and phase parameters, the direct estimator leaves the variational objective unchanged and estimates the same population phase force as the score estimator; only the finite-sample random variable changes. The adaptive mixture interpolates between the score and direct forms with a minimum-variance coefficient, so any deterministic convex mixture remains unbiased at the population level and the optimal coefficient minimizes raw phase-gradient variance. In training, this coefficient is estimated from the same finite batch inside the clipped, preconditioned pipeline; its behavior is assessed by the logged variance diagnostics and seed statistics.

The practical improvement is mostly reliability. Some standard-gradient seeds can be competitive, but the standard estimator produces more high-error or bad-basin runs; reducing this failure rate matters because seed restarts, learning-rate sweeps and architecture retuning are costly. The phase-network learning-rate sweeps in Fig.~\ref{fig:flux} and Extended Data Fig.~\ref{fig:ed_phaselr} should therefore be read as stress tests, not as a requirement for equally elaborate tuning of the new estimator. On lattice benchmarks the direct estimator has little measured per-step overhead because it reuses connected configurations already evaluated for the local energy. In the FQHE benchmark it costs more (about $2.8\times$ slower at the full walker batch; Methods), so the method partly converts reruns and hyperparameter search into a variance-aware estimator cost.

The numerical formulation used in the main lattice experiments deliberately uses separated amplitude and phase networks, where the unbiasedness identity is transparent because sampling does not depend on phase parameters. We then separate the broader claim into a raw-gradient extension for coupled real two-head networks and narrower single-network transfer tests in the shared-trunk ladder and FQHE benchmarks. In the coupled construction, the direct summand is a stopped-amplitude, active-phase derivative and must be combined with the amplitude/Pulay score term; the full pathwise derivative is not the VMC gradient. The covariance corrections for the adaptive coefficient are specified in Methods. Testing the resulting full-force variance reduction more broadly in shared representations, autoregressive states, electronic-structure and other fermionic continuum wavefunctions, and foundation-style quantum-state models is a natural next step. Quantum chemistry is a natural but architecture-dependent extension: the most direct targets are complex-valued VMC settings---magnetic fields, spin--orbit coupling, twisted or periodic boundary conditions, current-carrying states and Berry-phase problems---whereas real zero-field molecules would require a sign/nodal analogue rather than a literal phase estimator.

Overall, difficult NQS calculations should be diagnosed not only through architectures, symmetries, transferability and physical priors, but also through the stochastic estimator used to learn the phase. Pathwise derivatives, control variates and adaptive estimator combinations belong inside VMC, not only around it. For fermions, gauge fields, twisted boundaries, quantum Hall states, chiral interactions and real-time dynamics, the wavefunction-phase bottleneck is not only an architecture or capacity bottleneck; it is also a noise floor, and noise floors can be engineered down.

\section{Methods}

\subsection{Amplitude--phase parameterization}

For a discrete basis configuration $x$, we write a complex neural quantum state as
\begin{equation}
    \psi_\theta(x)=\exp[u_{\theta_1}(x)+\ii \phi_{\theta_2}(x)],
\end{equation}
where $u_{\theta_1}$ is the log-amplitude network and $\phi_{\theta_2}$ is the phase network. The local energy is
\begin{equation}
    \Eloc(x)=\sum_y H_{xy}\frac{\psi_\theta(y)}{\psi_\theta(x)}.
\end{equation}
The standard VMC gradient for a real parameter $\theta$ is
\begin{equation}
    \partial_\theta E
    =2\,\mathrm{Re}\left[
    \left\langle (\Eloc-E) O_\theta^*(x)\right\rangle_{|\psi|^2}
    \right],
\end{equation}
where $O_\theta(x)=\partial_\theta \log\psi_\theta(x)$. For the separated networks this gives
\begin{align}
    g_{\theta_1}^{\mathrm{amp}}
    &=2\left\langle [\mathrm{Re}\,\Eloc(x)-E]\,
    \partial_{\theta_1}u(x)\right\rangle,\\
    g_{\theta_2}^{\mathrm{std}}
    &=2\left\langle \mathrm{Im}\,\Eloc(x)\,
    \partial_{\theta_2}\phi(x)\right\rangle .
\end{align}

Two clarifications frame what follows. First, the standard phase estimator is unbiased: its expectation equals the exact phase force, so the difficulty studied in this work is statistical, not a formal bias. Concretely, resolving a gradient component from $N$ samples requires $N\gtrsim\mathrm{Var}(\hat g)/|g|^2$, and for the phase sector this ratio can grow without bound, because the force is a covariance that can become small through cancellation while the per-sample fluctuations of $\mathrm{Im}\,\Eloc$ remain large. Second, this noise problem is pre-asymptotic. At an exact eigenstate the local energy is constant over configurations, so $\mathrm{Im}\,\Eloc\equiv0$ pointwise and the standard phase estimator is then itself zero-variance, consistent with the zero-variance principle \cite{assaraf1999zero}; practical optimization operates far from that limit, and the crossover between the two regimes is made quantitative in the variance-mixing subsection below.

\subsection{Direct local-energy derivative}

The direct estimator differentiates the local energy at fixed sampled configuration:
\begin{equation}
    \partial_\theta \Eloc(x)
    =
    \sum_{y\ne x}H_{xy}\frac{\psi_\theta(y)}{\psi_\theta(x)}
    \left[O_\theta(y)-O_\theta(x)\right],
    \label{eq:direct_lattice}
\end{equation}
because the diagonal term $H_{xx}$ is independent of $\theta$. For phase parameters, $O_{\theta_2}(x)=\ii\,\partial_{\theta_2}\phi(x)$, so Eq.~\eqref{eq:direct_lattice} directly probes how changing the phase modifies the off-diagonal interference ratios.

The relation to the phase force follows from a Hermiticity identity that holds for any parameter and any basis. Differentiating $\Eloc=\psi^{-1}H\psi$ at fixed configuration and taking the $|\psi|^2$ expectation,
\begin{equation}
    \left\langle \partial_\theta \Eloc \right\rangle
    =\left\langle\left(\Eloc^{*}-\Eloc\right)O_\theta\right\rangle
    =-2\left\langle \mathrm{Im}\,\Eloc\;\ii O_\theta\right\rangle .
    \label{eq:hermiticity_identity}
\end{equation}
For separated phase parameters, $O_{\theta_2}=\ii\,\partial_{\theta_2}\phi$, so the expectation is purely real,
\begin{equation}
    \left\langle \partial_{\theta_2} \Eloc \right\rangle
    =2\left\langle \mathrm{Im}\,\Eloc\,
    \partial_{\theta_2}\phi\right\rangle ,
\end{equation}
and the direct local-energy derivative targets exactly the same phase force as the standard phase gradient, with a different Monte Carlo estimator.

Equation~\eqref{eq:hermiticity_identity} also delimits the scope of the method. For a parameter that affects the amplitude as well, $\partial_\theta u\neq 0$, the exact gradient decomposes as
\begin{equation}
    \partial_\theta E
    =2\left\langle\left[\mathrm{Re}\,\Eloc-E\right]\partial_\theta u\right\rangle
    +\mathrm{Re}\left\langle \partial_\theta \Eloc \right\rangle ,
    \label{eq:shared_decomposition}
\end{equation}
and the direct estimator alone omits the first (score) contribution: its bias relative to the full gradient is exactly $-g_\theta^{\mathrm{amp}}$. Thus a direct local-energy derivative by itself is not a valid full-gradient estimator for any parameter that also changes the amplitude. In the separated phase block this issue is absent because $\partial_{\theta_2}u=0$. In a coupled real two-head network the obstruction can instead be removed by restoring the amplitude/Pulay force and restricting the pathwise derivative to the phase output, with the amplitude values held fixed; this construction is derived in Sec.~\ref{sec:coupled_two_head}. A genuinely tied or holomorphic single-complex parameterization is not covered unless such a stopped-amplitude, active-phase derivative is a well-defined operation.

Two structural properties of the direct estimator follow immediately from Eq.~\eqref{eq:direct_lattice}.

\textbf{Diagonal decoupling.} The diagonal term drops out exactly: the $y=x$ ratio is identically one, so $\partial_\theta H_{xx}$-type contributions vanish sample by sample; in continuum notation the same statement reads $\partial_\theta V(\mathbf{R})=0$, so the direct estimator is sample-wise independent of diagonal (potential) fluctuations.

The diagonal-energy statement should be interpreted narrowly: the diagonal matrix element $H_{xx}$ drops out of the direct local-energy derivative in Eq.~\eqref{eq:direct_lattice}. For a purely separated phase network, a real diagonal term also does not directly enter the standard phase-gradient expectation, which depends on $\mathrm{Im}\,\Eloc$. The variance advantage is therefore more accurately stated as an estimator-level reduction observed in phase-structured systems. By contrast, in shared or single-complex-network parameterizations the score-function estimator couples every parameter to $\mathrm{Re}\,\Eloc$ fluctuations through Eq.~\eqref{eq:shared_decomposition}, and diagonal (potential) noise does leak into phase directions; that observation was the original motivation for this work.

\textbf{Global gauge invariance.}
The direct estimator depends on the phase network only through differences $\partial_{\theta_2}\phi(y)-\partial_{\theta_2}\phi(x)$ on connected pairs. Hence, global-phase directions vanish identically sample by sample. In an uncentered score-function implementation, the corresponding standard phase variable contains the pure-noise term $2\,\mathrm{Im}\,\Eloc\times\mathrm{const}$; a centered covariance implementation can remove this particular global mode. The broader variance comparison below therefore concerns the full standard/direct random variables, not only this gauge direction.

\subsection{Unbiased phase estimators and variance mixing}

The estimator comparison can be made explicit for each phase parameter $\beta$ and each sampled configuration $R$. Write
\begin{equation}
    \partial_\beta\log\Psi(R)=\ii\,\partial_\beta\Phi(R),
\end{equation}
and define two real per-configuration quantities
\begin{align}
    A(R)
    &=
    \mathrm{Im}\,\Eloc(R)\,\partial_\beta\Phi(R),
    \\
    B(R)
    &=
    \sum_{R'}
    \mathrm{Im}\!\left[
    H_{RR'}\frac{\Psi(R')}{\Psi(R)}
    \right]
    \partial_\beta\Phi(R').
\end{align}
The standard score estimator and the direct local-energy-derivative estimator can then be written as
\begin{equation}
    \hat g_\beta^{\mathrm{std}}(R)=2A(R),
    \qquad
    \hat g_\beta^{\mathrm{dir}}(R)=A(R)-B(R).
    \label{eq:ab_estimators}
\end{equation}
Hermiticity gives $\langle B\rangle=-\langle A\rangle$, so both estimators have the same expectation,
\begin{equation}
    \langle \hat g_\beta^{\mathrm{std}}\rangle
    =
    \langle \hat g_\beta^{\mathrm{dir}}\rangle
    =
    2\langle A\rangle .
    \label{eq:unbiased_equivalence}
\end{equation}
Thus the direct estimator does not change the exact variational gradient; it changes the finite-sample random variable used to estimate the same phase force.

Let
\begin{equation}
    a^2=\langle A^2\rangle,\qquad
    b^2=\langle B^2\rangle,\qquad
    c=\langle AB\rangle .
\end{equation}
Because the two estimators have the same mean, comparing their variances gives the simple criterion
\begin{equation}
    \mathrm{Var}\!\left(\hat g_\beta^{\mathrm{dir}}\right)
    <
    \mathrm{Var}\!\left(\hat g_\beta^{\mathrm{std}}\right)
    \quad\Longleftrightarrow\quad
    b^2-2c<3a^2 .
    \label{eq:variance_condition}
\end{equation}
This condition is not a universal inequality; it is an instance-dependent statement about the phase-gradient noise, and the two regimes of training sit on opposite sides of it.

\emph{Pre-asymptotic regime.}
The phase force is a covariance, $2\langle A\rangle=2\,\mathrm{Cov}(\mathrm{Im}\,\Eloc,\partial_\beta\Phi)$, since $\langle\mathrm{Im}\,\Eloc\rangle=0$ for Hermitian $H$. The force can therefore be small through cancellation while the noise floor $a^2=\langle A^2\rangle$ remains at the scale of $\langle(\mathrm{Im}\,\Eloc)^2\rangle$: the signal-to-noise ratio of the standard estimator can collapse long before the energy converges. This is the regime in which the direct estimator helps.

\emph{Eigenstate limit.}
At an exact eigenstate, $\mathrm{Im}\,\Eloc\equiv0$ pointwise, so $\hat g_\beta^{\mathrm{std}}$ is a zero-variance estimator. The direct estimator $\hat g_\beta^{\mathrm{dir}}=-B$ retains a generally nonzero variance, because only the unweighted sum over connected configurations vanishes at the eigenstate, not its $\partial_\beta\Phi(R')$-weighted counterpart. In this limit $V_s\to0$ and $C\to0$, so the optimal mixing coefficient defined below obeys $\lambda^\star\to0$: the adaptive-mixture returns to the standard estimator exactly when the zero-variance principle begins to favor it. The measured $\lambda^\star$ trajectories in Fig.~\ref{fig:variance}, which start near the direct endpoint and drift away from $\lambda^\star=1$ during training, follow this prediction.

The same decomposition suggests a variance-aware interpolation. The optimal convex combination of two unbiased estimators is a classical control-variate or minimum-variance linear-combination construction in Monte Carlo and machine-learning gradient estimation \cite{owen2013monte,greensmith2004variance,mohamed2020monte,tucker2017rebar,grathwohl2018backprop,ranganath2014black}. Our contribution is not this generic construction itself, but its use in the phase sector of NQS-VMC: we derive a direct local-energy-derivative phase estimator and apply the minimum-variance combination to the standard/direct phase-gradient pair. This is useful because the relative variance ordering of the two estimators can change during training, so a fixed choice of estimator is not obviously optimal. Since both endpoints are unbiased, any convex combination
\begin{equation}
    \hat g_{\beta,\lambda}
    =
    (1-\lambda)\hat g_\beta^{\mathrm{std}}
    +
    \lambda\hat g_\beta^{\mathrm{dir}},
    \qquad 0\le \lambda\le 1,
\end{equation}
targets the same population force for every deterministic $\lambda$. Denoting
\begin{equation}
    V_s=\mathrm{Var}(\hat g_\beta^{\mathrm{std}}),\quad
    V_d=\mathrm{Var}(\hat g_\beta^{\mathrm{dir}}),\quad
    C=\mathrm{Cov}(\hat g_\beta^{\mathrm{std}},\hat g_\beta^{\mathrm{dir}}),
\end{equation}
the unconstrained minimum-variance mixing coefficient is
\begin{equation}
    \lambda^\star
    =
    \frac{V_s-C}{V_s+V_d-2C},
    \label{eq:lambda_star}
\end{equation}
with the practical convex choice $\lambda^\star_{\mathrm{clip}}=\min[1,\max(0,\lambda^\star)]$.

The denominator is itself a variance, $V_s+V_d-2C=\mathrm{Var}(\hat g_\beta^{\mathrm{std}}-\hat g_\beta^{\mathrm{dir}})\ge0$, so the mixture variance $V(\lambda)\equiv\mathrm{Var}(\hat g_{\beta,\lambda})$ is a convex quadratic on $[0,1]$; convexity here is a theorem, not an assumption. (If the denominator vanishes, the two estimators differ by a constant and every $\lambda$ is equivalent.) Because the endpoints $\lambda=0,1$ lie in the feasible interval, the population minimum over convex mixtures obeys
\begin{equation}
    \mathrm{Var}\!\left(\hat g_{\beta,\lambda^\star_{\mathrm{clip}}}\right)
    \le
    \min\!\left[
    \mathrm{Var}(\hat g_\beta^{\mathrm{std}}),
    \mathrm{Var}(\hat g_\beta^{\mathrm{dir}})
    \right].
    \label{eq:mixing_population_opt}
\end{equation}
For an estimated coefficient, the same quadratic gives a sensitivity bound. Completing the square gives the exact identity
\begin{equation}
    V(\lambda)
    =V(\lambda^\star)
    +\left(V_s+V_d-2C\right)\left(\lambda-\lambda^\star\right)^2 ,
    \label{eq:lambda_degradation}
\end{equation}
so a deterministic or independently estimated coefficient whose value is $\hat\lambda$ has variance exceeding the population optimum by $(V_s+V_d-2C)\,(\hat\lambda-\lambda^\star)^2$ for that realization, and by $(V_s+V_d-2C)\,\mathbb{E}[(\hat\lambda-\lambda^\star)^2]$ after averaging over the coefficient estimate. Thus Eq.~\eqref{eq:mixing_population_opt} is an exact raw-population statement, while a finite coefficient estimate can deviate from the population optimum by an amount controlled by Eq.~\eqref{eq:lambda_degradation}. In production we use the corresponding trace version as a scalar plug-in coefficient,
\begin{equation}
    \lambda_t^\star
    =
    \mathrm{clip}_{[0,1]}
    \frac{\mathrm{Tr}\,V_s-\mathrm{Tr}\,C}
    {\mathrm{Tr}\,V_s+\mathrm{Tr}\,V_d-2\,\mathrm{Tr}\,C},
    \label{eq:trace_lambda}
\end{equation}
where the trace sums over phase-parameter components. The coefficient is estimated on the pre-update walker ensemble at the logged optimization state and then held fixed for the subsequent stochastic-reconfiguration update. No exponential moving average is used; the only stabilization is the convex clipping $\lambda_t^\star\in[0,1]$ and the same local-energy clipping used by the gradient estimators. In the separated-network experiments, both endpoints estimate the same phase force and the plug-in $\lambda_t^\star$ chooses the interpolation used for that step; any residual same-batch correlation between $\lambda_t^\star$ and estimator noise is a finite-sample implementation effect, not a change in the population force being targeted. We therefore monitor the actual finite-batch behavior directly through the logged $\lambda_t^\star$ and variance-ratio trajectories in Fig.~\ref{fig:variance}. With a single scalar coefficient, the population statement applies to the summed (trace) phase-gradient variance rather than to each component individually. The coefficient in Eq.~\eqref{eq:trace_lambda} is the one used for the separated amplitude--phase experiments. For a coupled two-head network the object whose variance is minimized is the full shared-parameter force, not the phase block alone; the corresponding coefficient contains amplitude--phase cross-covariances and is given in Sec.~\ref{sec:coupled_two_head}.

\subsection{Variance sets the attainable descent speed}

The practical consequence of estimator variance can be understood through an idealized unpreconditioned phase-sector SGD step; we restate the standard argument~\cite{bottou2018optimization} in the present setting. This subsection adapts the convergence analysis from an early version of this work and should be read as a regime statement under smoothness assumptions, not as a global convergence theorem for the nonconvex, preconditioned VMC dynamics used in production.

Assume the energy restricted to the phase sector (amplitude parameters held fixed at their current values) is $L$-smooth. For an unbiased estimator $\hat g$ of the sector gradient $\nabla E$, with $\mathbb{E}\|\hat g\|^2=\|\nabla E\|^2+V$, one stochastic-gradient step of size $\eta$ obeys
\begin{equation}
    \mathbb{E}[E_{t+1}]-E_t
    \le
    -\eta\|\nabla E\|^2
    +\tfrac{1}{2}L\eta^2\left(\|\nabla E\|^2+V\right).
\end{equation}
Optimizing this bound over $\eta$ gives $\eta^\star=\|\nabla E\|^2/[L(\|\nabla E\|^2+V)]$ and the maximal expected one-step decrease
\begin{equation}
    \Delta E_{\max}
    =
    -\frac{1}{2L}\,
    \frac{\|\nabla E\|^4}{\|\nabla E\|^2+V}.
    \label{eq:max_speed}
\end{equation}
Two regimes follow. When $V\ll\|\nabla E\|^2$, the decrease is the deterministic $-\|\nabla E\|^2/2L$. When $V\gg\|\nabla E\|^2$, it collapses to $-\|\nabla E\|^4/(2LV)$: a quartic stall in which progress is inversely proportional to the gradient variance. Combined with the pre-asymptotic signal-to-noise collapse identified above, this places noisy wavefunction-phase optimization in the stalled regime, while the amplitude sector --- whose residual $(\mathrm{Re}\,\Eloc-E)^2$ is suppressed by the variational principle itself \cite{assaraf1999zero} --- can descend at the deterministic rate.

Two corollaries connect Eq.~\eqref{eq:max_speed} to the experiments.
First, in this idealized unpreconditioned SGD model, if the optimizer is in the stalled regime $V \gg \|\nabla E\|^2$, a $\kappa$-fold reduction of the phase-gradient variance raises the optimized one-step descent bound by the same factor $\kappa$. The logged variance ratios in Fig.~\ref{fig:variance} should therefore be read as variance-controlled descent-speed headroom, rather than as an exact convergence theorem for the full minSR update.
Second, rescaling the phase-network learning rate cannot remove this raw-gradient variance ceiling: Eq.~\eqref{eq:max_speed} is already optimized over a scalar step size, so amplifying $\eta_\phi$ rescales signal and noise together, and steps beyond $\eta^\star$ violate the descent condition. This is consistent with the phase-network learning-rate baselines, which improve up to $\eta_\phi/\eta=3$, plateau above the direct estimator, and destabilize at large multipliers.

\subsection{Why the estimator can improve wavefunction-phase optimization}

The mechanism is quantitative at the estimator level: the $A/B$ decomposition gives the instance criterion of Eq.~\eqref{eq:variance_condition}, the diagonal and gauge structure of $g_{\rm dir}$ removes identifiable noise channels exactly, and Eq.~\eqref{eq:max_speed} explains how measured variance ratios translate into descent-speed headroom in an idealized raw-gradient model. We therefore formulate the mechanism as a conditional proposition rather than an unconditional convergence theorem: when the phase network is expressive enough, sampling is adequate, and the optimization does not become trapped in unrelated variational basins, replacing the standard phase-gradient estimator by the direct/adaptive-mixture local-energy derivative can reduce phase-gradient variance and accelerate descent.

\subsection{Separate amplitude and phase updates}

The numerical experiments in this work use separated amplitude and phase parameter blocks. In that setting, we keep the conventional amplitude update and replace only the phase estimator:
\begin{equation}
    \Delta\theta_1=-\eta_u\,g_{\theta_1}^{\mathrm{amp}},
    \qquad
    \Delta\theta_2=-\eta_\phi\,g_{\theta_2}^{\mathrm{dir}}.
\end{equation}
This avoids using the direct estimator as a biased amplitude update. At the level of exact population forces for a strictly separated amplitude--phase ansatz, the update combines the conventional amplitude force with a phase force whose expectation is identical for the standard, direct, and adaptive-mixture estimators. Because the two forces act on disjoint parameter blocks, the exact composite force vanishes if and only if both sector forces vanish. Thus, before finite-sample preconditioning, clipping, or same-batch estimates of $\lambda_t^\star$ are introduced, the stationary points of the sector-separated raw-gradient update coincide with those of the exact variational energy gradient; the estimator change does not introduce new population-level stationary points.

The estimator identities in this section concern raw, population gradient estimators. The production runs use the same minSR-style preconditioner for every compared method: amplitude and phase parameters are concatenated into one tangent-space system, with amplitude scores entering through the real log-amplitude derivatives and phase scores entering through the imaginary phase derivatives. The sampled configurations, MCMC schedule, damping, clipping, and linear-solve pipeline are identical for the standard, direct, and adaptive-mixture estimators; only the phase-force vector is changed. Because the preconditioner, the adaptive mixing coefficient, and the clipped force vectors are sample-dependent, the raw-gradient expectation equivalence is a population statement rather than a literal identity for every finite-sample preconditioned update. The theorem identifies the force targeted by the separated estimator, and the numerical comparisons test the complete preconditioned pipeline.

Two caveats bound the stationarity statement above, and they motivate the control that follows. First, the coincidence is with the stationary points of the parameter-space energy $E(\theta)$, which for a neural ansatz is weaker than reaching an eigenstate: the sector forces are the Hilbert-space gradient projected onto the network's tangent directions, so they can vanish at spurious critical points where the Hilbert-space gradient is nonzero but annihilated by the network Jacobian. No uniqueness or global-convergence claim is therefore made for arbitrary neural parameterizations. Second, the table ansatz over the full finite Hilbert space removes exactly this confounder: there every Hilbert-space direction is a parameter direction, the representation can express the exact eigenvector, and critical points on the full-support stratum are eigenstates, so table-ansatz experiments isolate estimator behavior from network expressivity. For neural networks the corresponding statement is empirical: across the benchmarks studied here, the direct phase estimator improves the reliability and the final energy whenever a non-trivial complex phase structure is present.

\subsection{Generalization to coupled two-head networks}
\label{sec:coupled_two_head}

The preceding estimator identities are used in this paper primarily for separated amplitude and phase networks. A closely related, but distinct, construction applies to a coupled real two-head ansatz
\begin{equation}
    \psi_\theta(x)=\exp[U_\theta(x)+\ii\Phi_\theta(x)],
    \qquad
    p_\theta(x)=\frac{e^{2U_\theta(x)}}{\sum_z e^{2U_\theta(z)}} ,
    \label{eq:coupled_ansatz}
\end{equation}
where the same real parameter vector $\theta$ can affect both $U_\theta$ and $\Phi_\theta$. This subsection gives the raw-gradient population identity for this case. It underlies the shared-trunk control in Extended Data Fig.~\ref{fig:ed_singlenet}, but the numerical coverage of coupled networks is much narrower than the separated-network benchmark suite.

Fix a current parameter value and an infinitesimal direction $v$. Write
\begin{equation}
    \dot U_v(x)=\nabla_\theta U_\theta(x)\!\cdot v,
    \qquad
    \dot\Phi_v(x)=\nabla_\theta\Phi_\theta(x)\!\cdot v,
\end{equation}
and denote $R(x)=\mathrm{Re}\,\Eloc(x)$ and $I(x)=\mathrm{Im}\,\Eloc(x)$ at the current $\theta$. Differentiating
\begin{equation}
    E(\theta)=\sum_x p_\theta(x)R_\theta(x)
\end{equation}
along $v$ gives
\begin{equation}
    D E(\theta)[v]
    =
    \sum_x \dot p(x)R(x)
    +
    \sum_x p_\theta(x)\dot R(x).
    \label{eq:coupled_product_rule}
\end{equation}
The first term is the derivative of the sampling law. Since $p_\theta\propto e^{2U_\theta}$,
\begin{equation}
    \dot p(x)
    =
    2p_\theta(x)\left[
        \dot U_v(x)-\langle\dot U_v\rangle
    \right],
\end{equation}
and therefore
\begin{equation}
    \sum_x\dot p(x)R(x)
    =
    2\,\mathrm{Cov}_{p_\theta}(R,\dot U_v).
    \label{eq:coupled_pulay}
\end{equation}
This is the amplitude/Pulay contribution. Its Monte Carlo summand is the state-dependent coefficient
\begin{equation}
    2\,[R(x_i)-\mu_R]\,\nabla_\theta U_\theta(x_i),
    \qquad
    \mu_R=\langle R\rangle ,
\end{equation}
so the sampling derivative is included precisely when the amplitude force is included.

The explicit local-energy derivative in Eq.~\eqref{eq:coupled_product_rule} splits as
\begin{equation}
    \dot R(x)
    =
    D_U R(x)[\dot U_v]
    +
    D_\Phi R(x)[\dot\Phi_v].
\end{equation}
For Hermitian, $\theta$-independent $H$, the explicit $U$-path term has zero real expectation. To see this, define
\begin{equation}
    B_{xy}
    =
    p_\theta(x)H_{xy}\frac{\psi_\theta(y)}{\psi_\theta(x)}
    =
    \frac{\psi_\theta(x)^*H_{xy}\psi_\theta(y)}
         {\sum_z|\psi_\theta(z)|^2}.
\end{equation}
Hermiticity gives $B_{yx}=\overline{B_{xy}}$. Hence
\begin{align}
    \left\langle D_U R[\dot U_v]\right\rangle
    &=
    \mathrm{Re}\sum_{x,y}B_{xy}\,[\dot U_v(y)-\dot U_v(x)] .
\end{align}
The contribution of the pair $(x,y)$ and $(y,x)$ is
\begin{equation}
    \mathrm{Re}\left[
      \bigl(B_{xy}-\overline{B_{xy}}\bigr)
      [\dot U_v(y)-\dot U_v(x)]
    \right]=0,
\end{equation}
because $B_{xy}-\overline{B_{xy}}$ is purely imaginary while $\dot U_v(y)-\dot U_v(x)$ is real. Thus
\begin{equation}
    \left\langle D_U R[\dot U_v]\right\rangle=0.
    \label{eq:coupled_U_cancel}
\end{equation}

The explicit phase term gives the usual phase force. Indeed,
\begin{equation}
    D_\Phi\Eloc(x)[\dot\Phi_v]
    =
    \ii\sum_y H_{xy}\frac{\psi_\theta(y)}{\psi_\theta(x)}
    [\dot\Phi_v(y)-\dot\Phi_v(x)].
\end{equation}
Using the same Hermitian pairing as above,
\begin{equation}
    \left\langle D_\Phi R[\dot\Phi_v]\right\rangle
    =
    2\left\langle I\,\dot\Phi_v\right\rangle
    =
    2\,\mathrm{Cov}_{p_\theta}(I,\dot\Phi_v),
    \label{eq:coupled_phase_identity}
\end{equation}
where the last equality uses $\langle I\rangle=0$ for Hermitian $H$.

Combining Eqs.~\eqref{eq:coupled_pulay}--\eqref{eq:coupled_phase_identity}, for every direction $v$,
\begin{equation}
    D E(\theta)[v]
    =
    2\,\mathrm{Cov}_{p_\theta}(R,\nabla_\theta U_\theta\!\cdot v)
    +
    \left\langle
      D_\Phi R[\nabla_\theta\Phi_\theta\!\cdot v]
    \right\rangle .
    \label{eq:coupled_directional}
\end{equation}
Equivalently, in vector form,
\begin{equation}
    \nabla_\theta E
    =
    g_{\mathrm{amp}}
    +
    g_{\mathrm{ph}},
    \qquad
    g_{\mathrm{amp}}
    =
    2\,\mathrm{Cov}_{p_\theta}(R,\nabla_\theta U_\theta),
    \qquad
    g_{\mathrm{ph}}
    =
    2\,\mathrm{Cov}_{p_\theta}(I,\nabla_\theta\Phi_\theta).
    \label{eq:coupled_vector_gradient}
\end{equation}
The phase force $g_{\mathrm{ph}}$ can equivalently be estimated by a stopped-amplitude pathwise derivative,
\begin{equation}
    g_{\mathrm{ph}}
    =
    \left\langle
    \nabla_\theta R(x)\big|_{\Phi\text{-path},\,U\text{ detached}}
    \right\rangle .
    \label{eq:coupled_direct_phase_population}
\end{equation}
Thus the coupled-network analogue of the direct estimator is not the bare pathwise derivative through all of $\theta$. It is the reconstructed force
\begin{equation}
    \widehat G_{\mathrm{dir}}^{\mathrm{coupled}}
    =
    \widehat A+\widehat D,
    \label{eq:coupled_direct_force}
\end{equation}
where
\begin{align}
    \widehat A
    &=
    \frac{2}{N}\sum_{i=1}^N
    [R_i-\mu_R]\,\nabla_\theta U_\theta(x_i),\\
    \widehat S
    &=
    \frac{2}{N}\sum_{i=1}^N
    [I_i-\mu_I]\,\nabla_\theta\Phi_\theta(x_i),\\
    \widehat D
    &=
    \frac{1}{N}\sum_{i=1}^N
    \nabla_\theta
    \mathrm{Re}\!\left[
      \sum_y H_{x_i y}
      \exp\!\left(
        \mathrm{sg}\,[U_\theta(y)-U_\theta(x_i)]
        +
        \ii[\Phi_\theta(y)-\Phi_\theta(x_i)]
      \right)
    \right].
    \label{eq:coupled_ASD}
\end{align}
Here $\mathrm{sg}$ denotes a stop-gradient operation in this summand. The standard coupled force is $\widehat A+\widehat S$, and the mixed coupled force is
\begin{equation}
    \widehat G_{\lambda}^{\mathrm{coupled}}
    =
    \widehat A+(1-\lambda)\widehat S+\lambda\widehat D .
    \label{eq:coupled_mix}
\end{equation}
With exact current-law sampling, fixed population baselines, and deterministic $\lambda$, Eqs.~\eqref{eq:coupled_directional}--\eqref{eq:coupled_mix} give
\begin{equation}
    \mathbb{E}\!\left[\widehat G_{\lambda}^{\mathrm{coupled}}\right]
    =
    \nabla_\theta E
    \qquad
    \text{for every fixed }0\le\lambda\le1.
\end{equation}
This is the raw-population identity. In a finite implementation using the adaptive version, the clipped coefficient $\lambda^\star_{\mathrm{coupled}}$ is estimated from samples and the force is also affected by empirical self-centering of $\mu_R,\mu_I$, stale Markov chains, clipping, and stochastic-reconfiguration preconditioning. These are implementation approximations around the same estimator-design principle; in coupled networks, where the amplitude and phase paths share parameters, they should be interpreted more cautiously than the separated-block identity.

Parameter sharing changes the variance optimization. For the separated phase block, the adaptive coefficient in Eq.~\eqref{eq:trace_lambda} minimizes the trace variance of the phase estimator alone. For a coupled network the full random vector is
\begin{equation}
    \widehat A+(1-\lambda)\widehat S+\lambda\widehat D .
\end{equation}
Let
\begin{equation}
    V_s=\mathrm{tr}\,\mathrm{Var}[\widehat S],
    \quad
    V_d=\mathrm{tr}\,\mathrm{Var}[\widehat D],
    \quad
    C=\mathrm{tr}\,\mathrm{Cov}[\widehat S,\widehat D],
    \quad
    A_s=\mathrm{tr}\,\mathrm{Cov}[\widehat A,\widehat S],
    \quad
    A_d=\mathrm{tr}\,\mathrm{Cov}[\widehat A,\widehat D].
\end{equation}
Then the trace-variance-minimizing unconstrained coefficient is
\begin{equation}
    \lambda^\star_{\mathrm{coupled}}
    =
    \frac{(V_s-C)+(A_s-A_d)}
         {V_s+V_d-2C},
    \label{eq:coupled_lambda}
\end{equation}
with the practical value clipped to $[0,1]$. The separated formula is recovered when $A_s=A_d$, in particular when the amplitude and phase forces have disjoint parameter supports. The new quantities $A_s$ and $A_d$ are the diagnostic signature of coupling: a direct phase estimator can reduce the phase-channel variance while still failing to reduce the full-force variance if its covariance with the amplitude force is unfavorable.

This coupled construction applies to real two-head networks for which the stopped-$U$, active-$\Phi$ derivative in Eq.~\eqref{eq:coupled_ASD} is well defined. It does not apply to a tied parameterization in which no such phase-path derivative can be formed. In particular, a genuinely holomorphic complex-weight network, where $U$ and $\Phi$ are Cauchy--Riemann tied as functions of the parameters, would require a separate analysis; mixing a biased full-pathwise derivative with the standard estimator would generally be biased for $\lambda>0$.

\subsection{Computational cost}

Per sampled configuration, the local energy already requires the wavefunction at every connected configuration $y$. In the separated implementation used for the reported runs, the direct estimator adds only the contraction of phase-network derivatives appearing in Eq.~\eqref{eq:direct_lattice} with the corresponding off-diagonal Hamiltonian ratios. This contraction can be evaluated as a batched vector-Jacobian product of the phase network over the $(\mathrm{batch}\times\mathrm{connected})$ configuration tensor, using the same weights that enter the local-energy derivative; it does not require materializing the full per-sample Jacobian $\partial_{\theta_2}\phi(y)$. On the lattice this uses first-order logarithmic derivatives only, with no spatial Laplacians. Operationally, the separated direct estimator requires one additional reverse-mode/VJP evaluation over the already-computed off-diagonal local-energy graph, whose size is set by the Hamiltonian connectivity, and it reuses the $\psi(y)$ forward evaluations while leaving the amplitude network untouched. In the coupled two-head extension of Sec.~\ref{sec:coupled_two_head}, the $U$ values in the direct summand are detached rather than differentiated, while the shared trunk still receives gradients through the $\Phi$ output. Its cost should therefore be measured separately for the given architecture; the population identity does not by itself guarantee the same wall-time overhead as the separated implementation. The archived A100 wall-time logs for the separated $L=50$ flux-ladder setting, batch size 1024, and 48 MCMC sweeps per step show no resolvable per-step overhead at this scale: after excluding the initial compilation/transient region, the median logged time per optimization step is 0.142 s for the standard estimator, 0.144 s for the direct estimator, and 0.133 s for the adaptive mixture. We therefore interpret the reported training-step comparisons as essentially per-unit-time comparisons within run-to-run timing noise, rather than as hiding a large computational-cost penalty. This near-free overhead is specific to the lattice, where the phase force needs only first-order logarithmic derivatives over the finite set of connected configurations. On the continuous-space fractional quantum Hall benchmark the direct estimator instead carries a genuine cost: the local energy contains the second-order Laplacian kinetic term, and the direct estimator differentiates it through the phase network for every walker, materializing a grad-through-Laplacian graph over the full walker batch rather than reusing a precomputed off-diagonal graph. Measured on a single A100 ($40$~GB) at $N=6$, $\nu=1/3$, batch size $1120$, in matched compile-excluded steady state, the standard estimator runs at $0.108$~s per optimization step with $5.0$~GB peak GPU memory, whereas the full direct estimator over all $1120$ walkers runs at $0.300$~s per step ($2.8\times$ slower) with $36$~GB peak memory ($+31$~GB), nearly saturating the device. Subsampling the direct-gradient contraction to $280$ walkers recovers most of this---$0.167$~s per step ($1.5\times$) and $13$~GB peak ($+8$~GB over standard)---and is the setting used for the reported FQHE runs, with no resolvable change in the reported energies. The continuum estimator overhead therefore scales with the walker batch and is controlled by this subsample.

\subsection{Idealized descent path in the table-ansatz limit}

It is useful to separate two statements that are often conflated. In the infinite-sample limit, $g_{\theta_2}^{\mathrm{dir}}$ and the standard phase gradient have the same expectation for separated phase parameters; therefore the direct estimator does not define a different variational energy surface. Its advantage is the estimator-level change in noise and conditioning analyzed above. A stronger convergence statement is available only in an idealized full-Hilbert-space limit, under assumptions that are standard in the nonconvex-optimization literature but that we do not prove for the sampled dynamics.

Consider a table ansatz on a finite Hilbert space, with independent log-amplitude and phase parameters $(u_x,\phi_x)$ for every basis configuration, and let $p(x)=|\psi(x)|^2/\sum_z|\psi(z)|^2$. The exact sector forces coincide with the coordinate partial derivatives,
\begin{equation}
    \frac{\partial E}{\partial u_x}=2\,p(x)\left[\mathrm{Re}\,\Eloc(x)-E\right],
    \qquad
    \frac{\partial E}{\partial \phi_x}=2\,p(x)\,\mathrm{Im}\,\Eloc(x),
    \label{eq:table_forces}
\end{equation}
so the exact separated update
\begin{equation}
    \dot{\theta}_1=-g_{\theta_1}^{\mathrm{amp}},
    \qquad
    \dot{\theta}_2=-g_{\theta_2}^{\mathrm{dir}}
\end{equation}
is a genuine gradient flow of the Rayleigh quotient $E[\psi]=\langle\psi|H|\psi\rangle/\langle\psi|\psi\rangle$ in amplitude--phase coordinates, with
\begin{equation}
    \frac{dE}{dt}
    =
    -\left\|\nabla_\theta E\right\|^2
    \le 0 .
    \label{eq:ideal_descent}
\end{equation}
Relative to the Hilbert-space metric, the coordinate flow is preconditioned by a positive diagonal weight proportional to $p(x)$, which degenerates as $p(x)\to0$.

On the full-support stratum, the stationary points of the Rayleigh quotient are eigenstates up to gauge, and every excited eigenstate is a strict saddle: if $|n\rangle$ has energy $E_n$ and $|m\rangle$ is any lower-energy eigenstate, then
\begin{equation}
    |\psi(\epsilon)\rangle
    =
    \sqrt{1-\epsilon^2}\,|n\rangle+\epsilon |m\rangle
    \quad\Rightarrow\quad
    E[\psi(\epsilon)]
    =
    E_n+\epsilon^2(E_m-E_n),
\end{equation}
exactly, which decreases for $E_m<E_n$. Convergence of the idealized flow to the ground-state manifold then rests on three assumptions: (A1) exact expectations replace Monte Carlo estimates; (A2) the support stays bounded away from zero along the trajectory, so the $p(x)$ weight never degenerates; and (A3) generic initializations avoid the measure-zero strict-saddle manifolds, as established for gradient methods on smooth nonconvex landscapes \cite{lee2016gradient}, with convergence of the iterates for analytic costs following from \L{}ojasiewicz-type arguments \cite{absil2005convergence}; Rayleigh-quotient flows are a classical instance of this theory \cite{helmke1994optimization}. These are standard assumptions rather than statements we prove for the sampled dynamics.

We emphasize that (A2) is a genuine restriction, not a technicality: by Eq.~\eqref{eq:table_forces} the force on $u_x$ is proportional to $p(x)$, so configurations whose weight collapses stop receiving updates, and the flow possesses additional quasi-stationary states at the support boundary that are not eigenstates of the full Hamiltonian. This boundary failure mode --- distinct from interior saddles --- is the idealized counterpart of the bad-basin trajectories observed in finite-sample runs, and it is not removed by either phase estimator.

For neural networks this picture becomes an assumption rather than a theorem: the network image must contain a lift of a descending trajectory, and the parameter landscape must not introduce additional spurious local minima. The role of the numerical experiments is precisely to test how well this idealized estimator-level picture survives in finite-width neural networks and finite-sample VMC.

\subsection{Flux-ladder Hamiltonian}

The primary benchmark is a two-leg ladder with complex Peierls phases on the leg exchange terms. In the spin representation,
\begin{align}
H_{\mathrm{ladder}}
=&
\sum_{\langle ij\rangle\in \mathrm{legs}}
\left[
J_z S_i^z S_j^z
+
\frac{J_{\mathrm{leg}}}{2}
\left(e^{\ii A_{ij}}S_i^+S_j^-+\mathrm{h.c.}\right)
\right]\nonumber\\
&+\sum_{\langle ij\rangle\in \mathrm{rungs}}
\left[
J_z S_i^z S_j^z
+
\frac{J_{\mathrm{rung}}}{2}
\left(S_i^+S_j^-+\mathrm{h.c.}\right)
\right],
\label{eq:flux_ladder_ham}
\end{align}
where $A_{ij}=+\Phi/2$ on one leg and $A_{ij}=-\Phi/2$ on the other leg in the symmetric gauge. We use $J_{\mathrm{leg}}=1$, $J_{\mathrm{rung}}=0.8$, and $J_z=0.5$. The main benchmark uses $L=50$ rungs in the fixed $S^z_{\mathrm{tot}}=0$ sector, equivalently $N_\uparrow=N/2$ for the spin basis. The legs have open boundary conditions, and the finite DMRG references use the same open ladder Hamiltonian and rung-by-rung site ordering. The references were generated with a TeNPy finite-MPS DMRG workflow using conservation of total magnetization, an MPO built from the same lattice edges, maximum bond dimension up to $\chi=384$ for the main $L=50$ references, singular-value cutoff $10^{-10}$, and up to 20 sweeps; the resulting finite-system reference energies used here are $\Eref=-44.826$ at $\Phi=0.2\pi$ and $\Eref=-43.303$ at $\Phi=0.3\pi$.

\subsection{Interacting spinless-fermion flux ladder}

The interacting-fermion Extended Data benchmark uses a 100-site spinless-fermion two-leg flux ladder at half filling, with nearest-neighbour interaction $V=2$ and plaquette flux $\Phi=0.5\pi$. In the gauge convention used for the spin ladder, the corresponding fermion Hamiltonian is written as
\begin{align}
H_{\mathrm{f}}
=&-t_{\mathrm{leg}}
\sum_{\langle ij\rangle\in\mathrm{legs}}
\left(e^{\ii A_{ij}}c_i^\dagger c_j+\mathrm{h.c.}\right)
-t_{\mathrm{rung}}
\sum_{\langle ij\rangle\in\mathrm{rungs}}
\left(c_i^\dagger c_j+\mathrm{h.c.}\right)\nonumber\\
&+V\sum_{\langle ij\rangle}(n_i-\tfrac12)(n_j-\tfrac12),
\label{eq:fermion_ladder_ham}
\end{align}
where $n_i=c_i^\dagger c_i$, $A_{ij}=+\Phi/2$ on one leg and $A_{ij}=-\Phi/2$ on the other, and $t_{\mathrm{leg}}=t_{\mathrm{rung}}=1$. The density interaction is written in the particle-hole-symmetric form used in the production runs (the interaction enters the Jordan--Wigner-mapped spin Hamiltonian as a uniform $J_z=V$ bond term, with no on-site shift). The model is represented in the fixed-particle-number sector and Jordan--Wigner mapped to a spin basis for the NQS implementation. The finite-system reference energy used for the plotted relative errors is obtained from a high-bond-dimension finite-MPS DMRG calculation with the same Hamiltonian, boundary geometry, and particle-number sector. Convergence with bond dimension and the final reference value are recorded in the data-audit table for Extended Data Fig.~\ref{fig:ed_fermion}; this benchmark is used as an additional generality check for sign-structured fermionic phases.

\subsection{Chiral XXX benchmarks and controls}

The chiral XXX benchmarks follow the time-reversal-breaking chiral spin-chain setting discussed in Ref.~\cite{wei2024unveiling}. The working model is a spin-$1/2$ XXX chain supplemented by a scalar-chirality interaction,
\begin{equation}
H_{\mathrm{chiral}}
=J\sum_i \mathbf S_i\cdot \mathbf S_{i+1}
+\alpha \sum_i \mathbf S_i\cdot(\mathbf S_{i+1}\times \mathbf S_{i+2}),
\label{eq:chiral_ham}
\end{equation}
\noindent We use periodic boundary conditions and isotropic exchange $J=1$ throughout. The small-system references ($N\le24$) are obtained by exact diagonalization in the fixed-magnetization sector, with the three-spin scalar-chirality term included exactly, while the larger sizes ($N=50$ and $N=100$) use finite-MPS DMRG references. The reference values used to compute the plotted errors include $\Eref=-9.5149$ for $N=20$, $\alpha=1.0$, $\Eref=-11.378061$ for $N=24$, $\Eref=-23.748041$ for $N=50$, and $\Eref=-47.469523$ for $N=100$. Relative energy error is reported as $|E-\Eref|/|\Eref|$, and the data-audit tables list the source paths and seeds. We scan $\alpha$, system size, and optimizer/ansatz controls.

\subsection{Two-dimensional flux-square benchmark}

The two-dimensional benchmark is an $8\times8$ ($64$-site) square lattice of spin-$1/2$ moments with a uniform Peierls flux $\Phi$ per plaquette. The Hamiltonian is
\begin{align}
H_{\square}
= \sum_{\langle ij\rangle}
\left[
J_z S_i^z S_j^z
+
\frac{J_{xy}}{2}
\left(e^{\ii A_{ij}}S_i^+S_j^-+\mathrm{h.c.}\right)
\right].
\label{eq:square2d_ham}
\end{align}
We use the Landau gauge $A_{(x,y),(x+1,y)}=-\Phi y$ and $A_{(x,y),(x,y+1)}=0$, so every plaquette encloses flux $\Phi$. We use $J_{xy}=1$ and, holding the cylinder geometry (periodic $x$, open $y$) fixed, scan the flux $\Phi\in\{0.2\pi,0.3\pi\}$ and the interaction $J_z\in\{0,0.5\}$ in the fixed $S^z_{\mathrm{tot}}=0$ sector, giving four cells. NQS runs use ten seeds per estimator with the same sampler and minSR pipeline as the flux ladder, but a wider MLP (width $256$, depth $2$; the one-dimensional ladders use width $128$, depth $2$). DMRG references were computed with TeNPy finite-MPS DMRG using total-magnetization conservation and singular-value cutoff $10^{-10}$, with a bond-dimension ladder up to $\chi=768$ to control the finite-bond-dimension error. The per-cell reference energies are $\Eref=-28.6722$ ($\Phi=0.2\pi$, $J_z=0$), $-32.7913$ ($\Phi=0.2\pi$, $J_z=0.5$), $-27.0269$ ($\Phi=0.3\pi$, $J_z=0$), and $-31.7306$ ($\Phi=0.3\pi$, $J_z=0.5$); relative energy errors are computed against these and are listed per seed in the data-audit table \texttt{square2d\_cylinder\_tail30\_summary.csv}.

\subsection{Neural architectures and optimization}

Unless otherwise stated, flux-ladder neural runs use separate real networks for log-amplitude and phase; the exception is the shared-trunk two-head control in Extended Data Fig.~\ref{fig:ed_singlenet}. The main MLP has width 128 and depth 2. Additional sweeps use wider/deeper MLPs and a convolutional ResNet for the phase: translation-equivariant residual convolutional blocks ($24$ channels, depth $6$) followed by spatial sum-pooling over the lattice, with a fixed sinusoidal positional encoding of the rung index (so the architecture is translation-equivariant but not exactly translation-invariant, and is not a full translation-group symmetrization). Training uses stochastic reconfiguration/minSR-style updates with identical hyperparameters across compared estimators unless specified in a sweep. The main runs use batch size $1024$, $100$ warmup sweeps, $48$ MCMC sweeps per optimization step, base learning rate $\eta=0.03$, learning-rate decay factor $0.3$, diagonal shift $0.02$, minimum diagonal shift $0.005$, and update-norm clipping at $5.0$. The minSR solve concatenates amplitude and phase parameters into a single preconditioned system $G=MM^\mathsf{T}+\lambda_{\mathrm{diag}}I$, then applies the same damping and solve to every estimator. The main sampler uses global fixed-magnetization exchange proposals: one up spin and one down spin are drawn and swapped, preserving $S^z_{\mathrm{tot}}=0$. The logged acceptance is the mean Metropolis acceptance over the proposal sweeps at a training step. The ResNet comparison uses ten seeds per estimator. Exceptions to the otherwise-matched hyperparameters are specified in the corresponding figure captions: the ResNet comparison uses batch size $512$ for the standard-gradient runs and $256$ for the direct runs (an asymmetry favoring the standard baseline), and the shared-trunk two-head control (Extended Data Fig.~\ref{fig:ed_singlenet}) uses a smaller learning rate ($\eta=0.01$) and diagonal floor ($\gamma_{\mathrm{min}}=0.0017$).

The same estimator-independent training-stabilization safeguard is enabled for every compared flux-ladder method. It is used only to keep stochastic optimization from catastrophic numerical excursions and is applied identically across estimators; the supplementary seed-level tables retain the corresponding event counts. Reported final bar plots use the tail-30 logged energy window for all displayed methods. Reported curves aggregate the seed traces using mean, median, and interquartile ranges. For the flux-ladder benchmarks all ten direct-gradient and adaptive-mixture seed traces are included, and standard-gradient phase-network learning-rate baselines are reported using the same seed convention in the processed source tables.

\subsection{Variance diagnostics}

For variance diagnostics, the standard, direct, and adaptive-mixture phase-gradient estimators are evaluated at logged adaptive-mixture training states using the same sampled configurations. We report the trace variance, i.e. the sum over phase-parameter components of the sample variance, and ratios of this trace variance between the standard estimator and the comparison estimator. These logged diagnostics correspond to the separated phase block used in the reported experiments. For a coupled two-head implementation, the analogous diagnostics should also include the full-force trace variances of $\widehat A+\widehat S$, $\widehat A+\widehat D$, and $\widehat A+(1-\lambda^\star_{\mathrm{coupled}})\widehat S+\lambda^\star_{\mathrm{coupled}}\widehat D$, together with the cross-covariances $A_s$ and $A_d$ in Eq.~\eqref{eq:coupled_lambda}.

\section*{Data availability}

The processed seed-level tables, chiral XXX JSON summary, the 2D flux-square per-seed table, and the per-cell 2D DMRG reference energies ($\chi$ up to 768) are included in the submission package \texttt{tables/} directory, together with \texttt{DATA\_AUDIT.md} and \texttt{SEEDS\_AND\_SOURCES.md}. The processed data and plotting scripts used to generate the figures will be made available in the project repository upon publication.

\section*{Code availability}

The code implements separated amplitude and phase networks, standard and direct/adaptive-mixture phase-gradient estimators, stochastic reconfiguration/minSR updates, and benchmark scripts for the flux ladder, chiral XXX, and 2D flux-square models. A public release will accompany the manuscript.

\section*{Acknowledgements}

We acknowledge Yubing Qian, Xiaoyong Ni and Khachatur Nazaryan for fruitful discussions. We also thank Ji Chen's group for making the DeepHall code available, parts of which served as a reference for our implementation. This work was supported by the National R\&D Program of China (2024YFA1410500, 2022YFA1403601), the Innovation Program for Quantum Science and Technology (Grant No. 2021ZD0302800), the National Natural Science Foundation of China (No. 12322402, No. 12274206), the Natural Science Foundation of Jiangsu Province (No. BK20233001), the Fundamental Research Funds for the Central Universities (No. KG202501), the Armenian Higher Education and Science Committee ARPI Remote Laboratory program 24RL-1C024, research projects 21AG-1C024 and 25Post-Doc1C003.

\section*{Competing interests}

The authors declare no competing interests.

% \bibliography{references}

\clearpage
\section*{Extended Data}
\setcounter{figure}{0}
\renewcommand{\figurename}{Extended Data Fig.}

\begin{figure}[t]
\centering
\includegraphics[width=\linewidth]{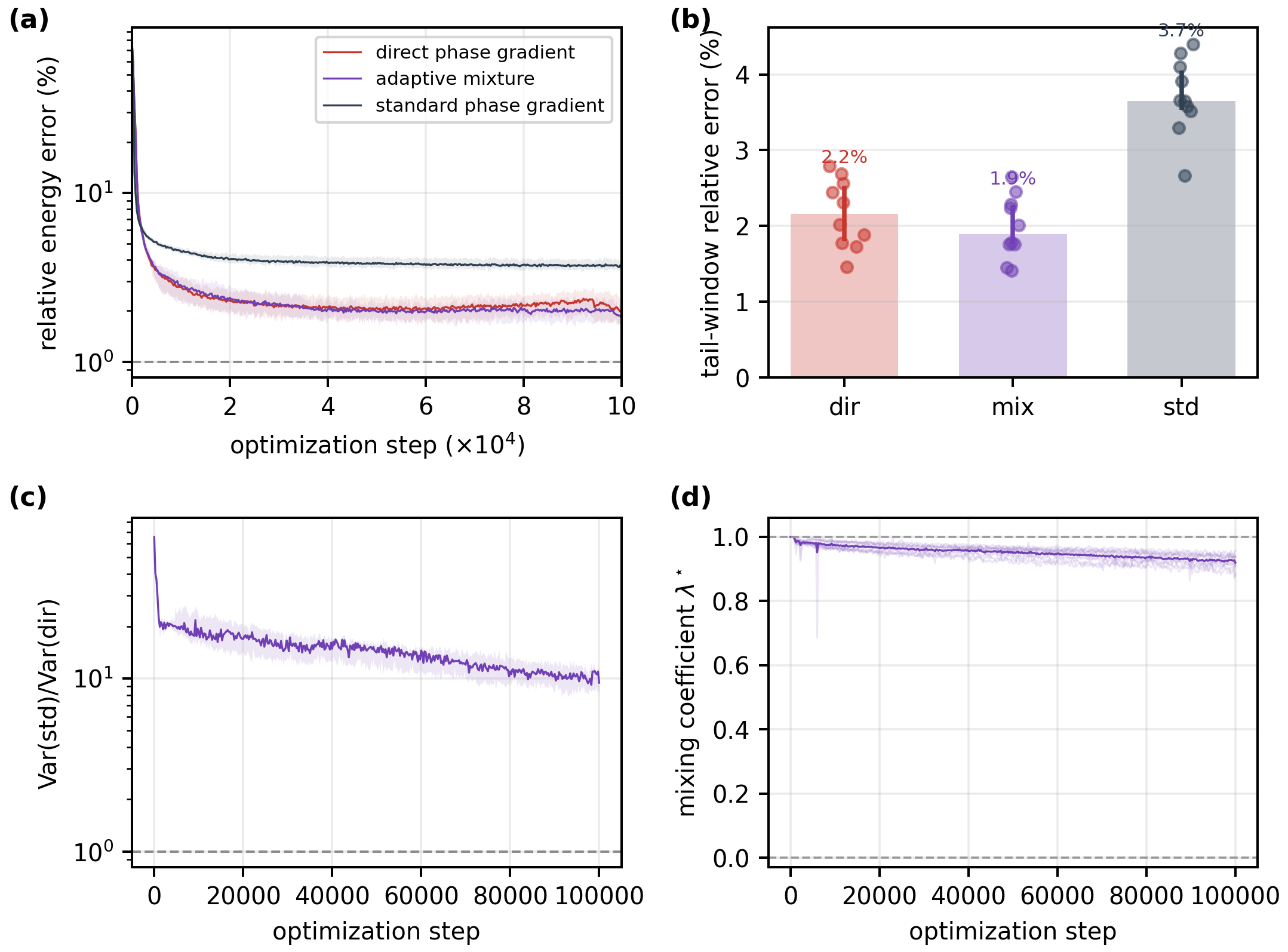}
\caption{\textbf{Generality to interacting fermions.}
A 100-site spinless-fermion two-leg flux ladder (nearest-neighbour interaction $V=2$, $\Phi=0.5\pi$, ten seeds), Jordan--Wigner mapped to spins: relative-error training curves, tail-window bars, the standard/direct variance ratio, and the mixing coefficient $\lambda^\star$. Against the DMRG reference $\Eref=-106.957$, the median tail-window errors are $2.2\%$ (direct) and $1.9\%$ (adaptive mixture) versus $3.7\%$ (standard); the standard/direct gradient-variance ratio exceeds an order of magnitude throughout training.}
\label{fig:ed_fermion}
\end{figure}

\begin{figure}[t]
\centering
\includegraphics[width=\linewidth]{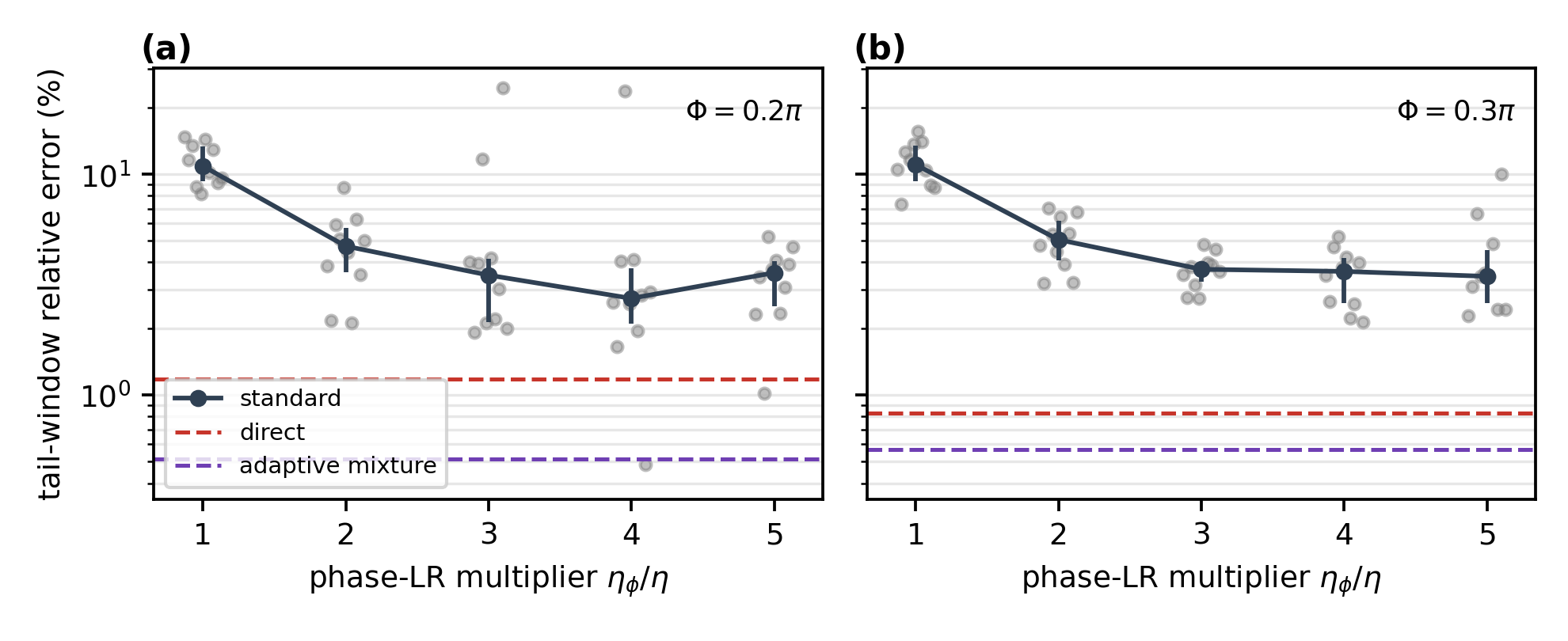}
\caption{\textbf{Full phase-learning-rate sweep.}
Standard phase-gradient baselines at multipliers $\eta_\phi/\eta=1$--$5$ on the 100-site flux ladder; the standard estimator plateaus well above the direct and adaptive-mixture estimators at every multiplier.}
\label{fig:ed_phaselr}
\end{figure}

\begin{figure}[t]
\centering
\includegraphics[width=\linewidth]{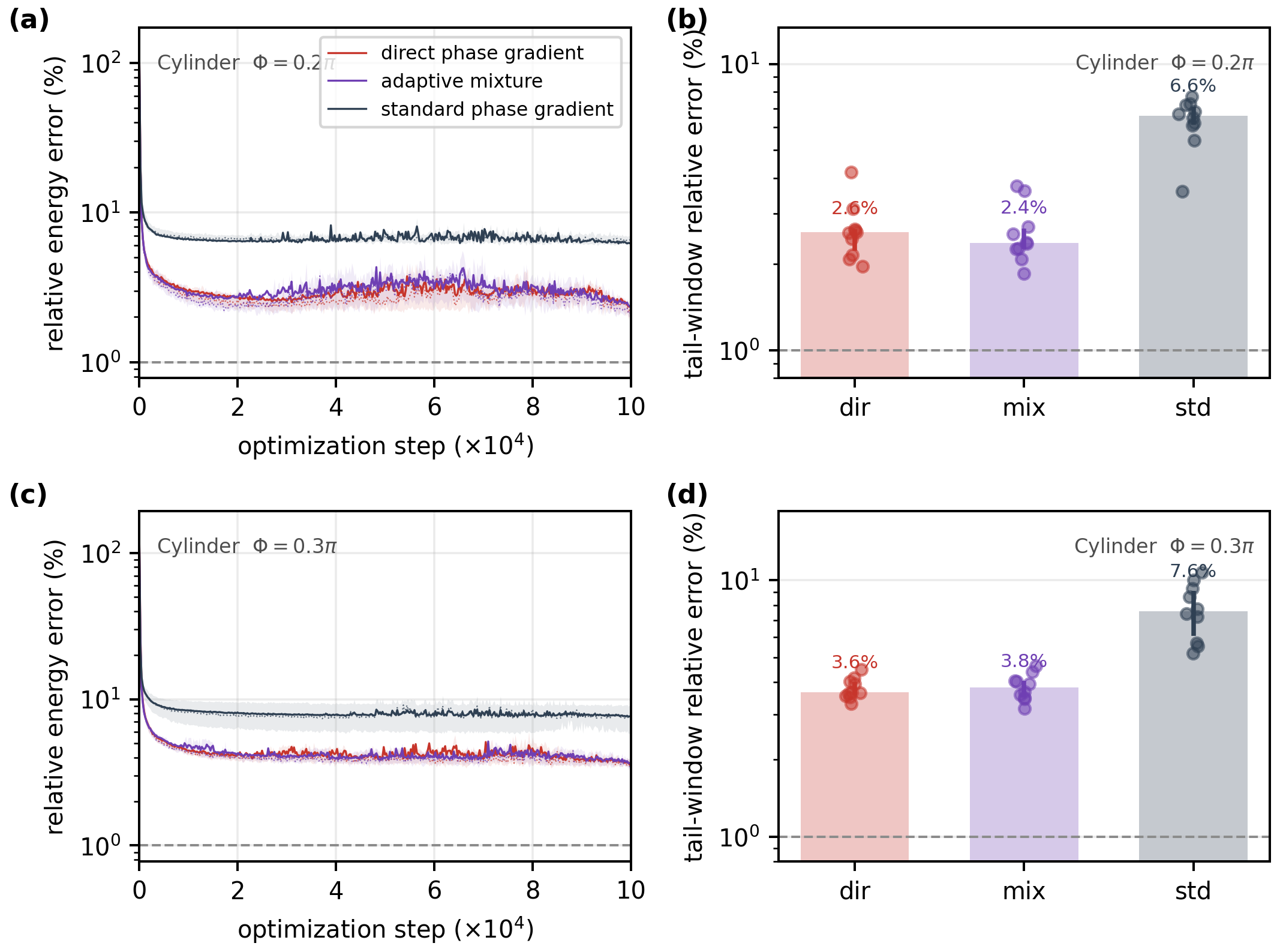}
\caption{\textbf{2D flux cylinder at the interacting point $J_z=0.5$.}
Relative-error training curves (left column) and tail-window relative-error statistics (right column; bars, median; whisker, interquartile range; points, individual seeds) on the $8\times8$ cylinder at $J_z=0.5$ (the main text shows $J_z=0$), for $\Phi=0.2\pi$ (top row) and $\Phi=0.3\pi$ (bottom row). In the training panels, solid lines show means, dotted lines medians, and shaded bands interquartile ranges; ten seeds per estimator, $10^5$ optimization steps.}
\label{fig:ed_2d}
\end{figure}

\begin{figure}[t]
\centering
\includegraphics[width=0.62\linewidth]{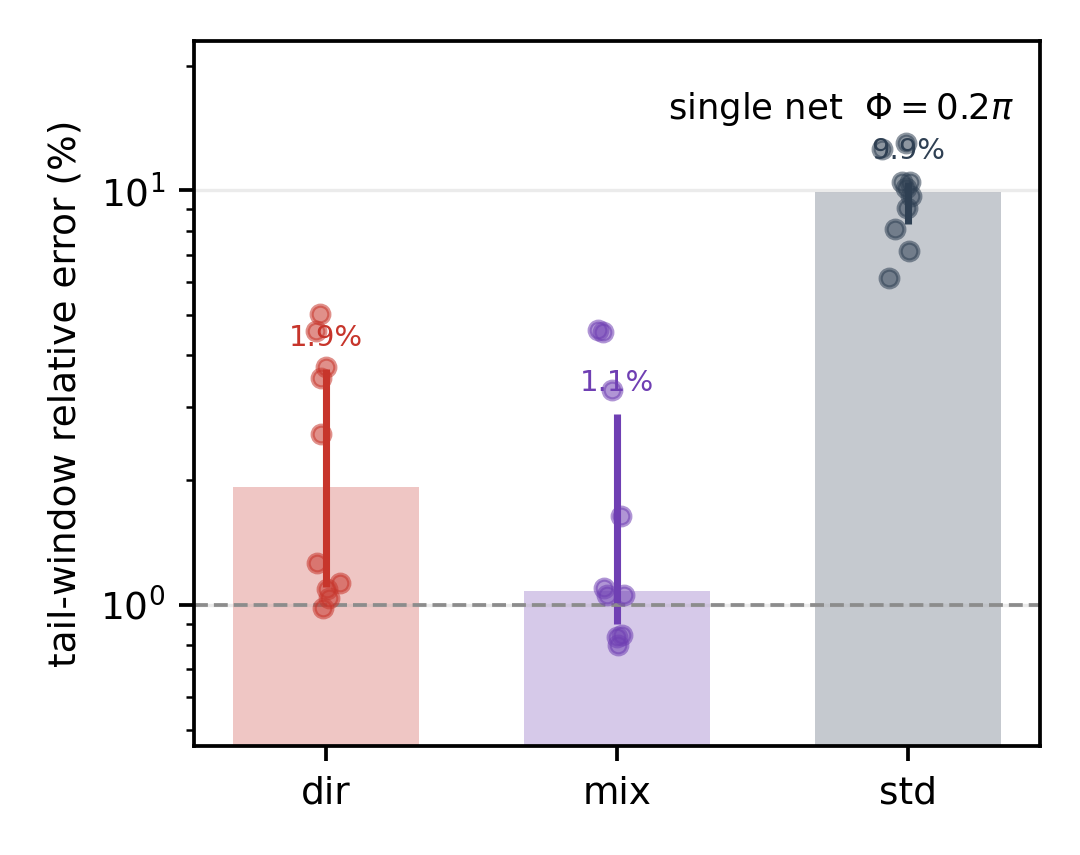}
\caption{\textbf{Single shared-network flux ladder at $\Phi=0.2\pi$.}
The main flux-ladder runs use separate amplitude and phase networks; here a single shared-trunk MLP (width $256$, depth $2$, tanh activations) feeds two linear heads that output the log amplitude and the phase. The shared trunk is widened to $256$. These runs use the same minSR pipeline as the main flux-ladder runs but a smaller learning rate ($\eta=0.01$) and diagonal floor ($\lambda_{\mathrm{min}}=0.0017$). For the direct and adaptive-mixture estimators, the coupled-force construction retains the amplitude/Pulay score term and applies the direct derivative only along the stopped-amplitude phase path. Median tail-window relative error (bars; whisker, interquartile range; tail $=$ last 30 logged points) per estimator with individual-seed points (ten seeds) on the $L=50$ ladder ($\Eref=-44.826$, DMRG). The direct phase-gradient ($1.92\%$) and adaptive-mixture ($1.08\%$) estimators remain several times more accurate than the standard phase-gradient estimator ($9.92\%$), so the advantage is not an artifact of splitting amplitude and phase into separate networks.}
\label{fig:ed_singlenet}
\end{figure}

\begin{figure}[t]
\centering
\includegraphics[width=\linewidth]{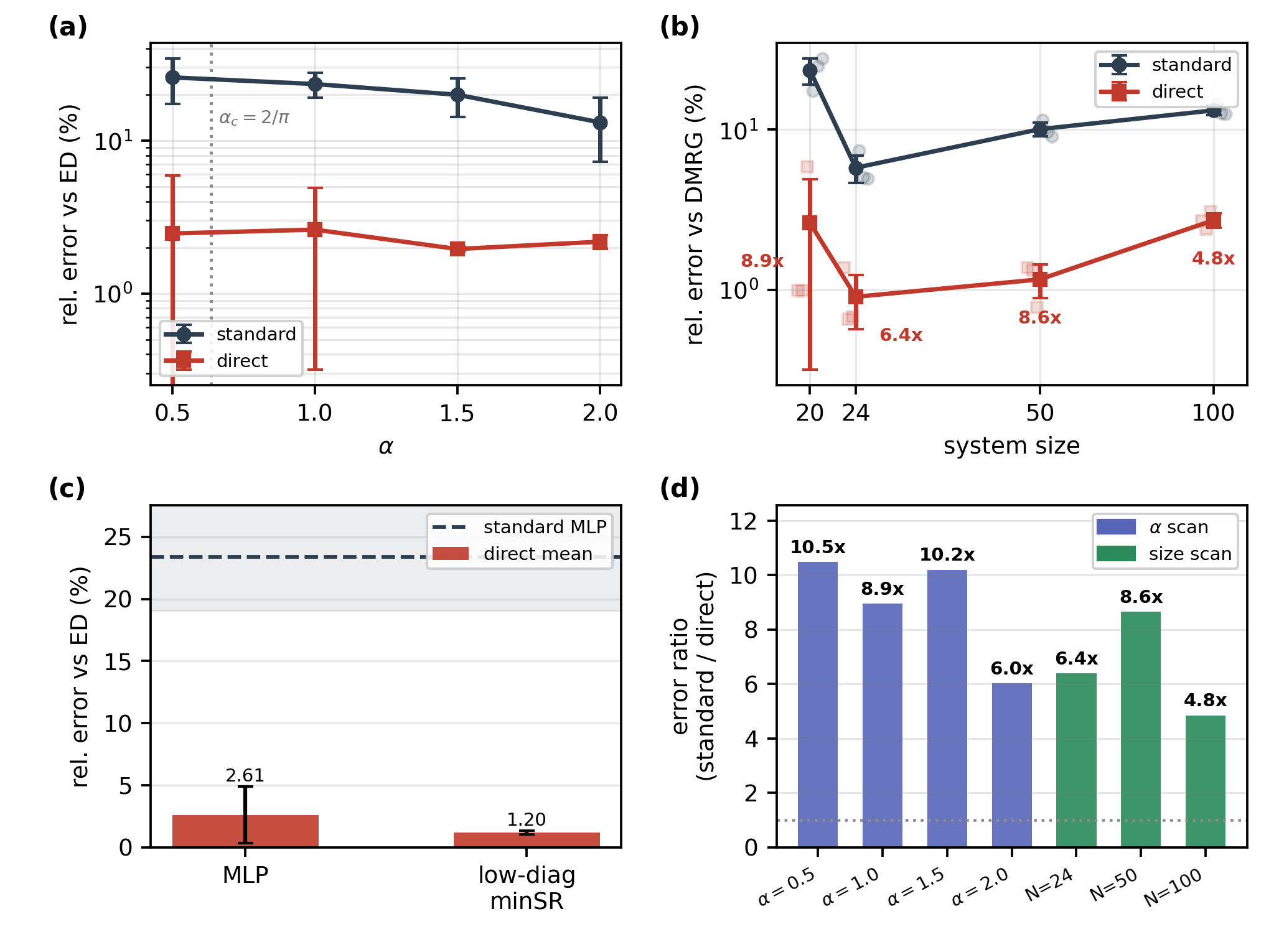}
\caption{\textbf{Chiral XXX benchmarks.}
(a) $\alpha$ scan at fixed size $N=20$ over $\alpha=0.5,1.0,1.5,2.0$.
(b) Size scaling at $\alpha=1.0$ for $N=20,24,50,100$.
(c) Optimizer control at $\alpha=1.0$, comparing the default MLP/minSR setting with a lower-diagonal minSR setting.
(d) Summary of the standard/direct error ratios from the $\alpha$ and size scans.}
\label{fig:ed_chiral}
\end{figure}

\end{document}